\pdfoutput=1
\documentclass{ecai} 

\usepackage{latexsym}
\usepackage{amssymb}
\usepackage{amsmath}
\usepackage{amsthm}
\usepackage{booktabs}
\usepackage{enumitem}
\usepackage{graphicx}
\usepackage{color}
\usepackage{multirow}
\usepackage{colortbl}
\usepackage{xcolor}

\newcommand{\BibTeX}{B\kern-.05em{\sc i\kern-.025em b}\kern-.08em\TeX}

\begin{document}


\begin{frontmatter}


\paperid{1228} 


\title{Anatomical Consistency Distillation and \\Inconsistency Synthesis for Brain Tumor Segmentation \\with Missing Modalities}


\author[A]{\fnms{Zheyu}~\snm{Zhang}}
\author[A]{\fnms{Xinzhao}~\snm{Liu}}
\author[A]{\fnms{Zheng}~\snm{Chen}}
\author[A]{\fnms{Yueyi}~\snm{Zhang}}
\author[C]{\fnms{Huanjing}~\snm{Yue}}
\author[D]{\fnms{Yunwei}~\snm{Ou}}
\author[A,B]{\fnms{Xiaoyan}~\snm{Sun}\thanks{Corresponding Author. Email: sunxiaoyan@ustc.edu.cn}}

\address[A]{MoE Key Laboratory of Brain-inspired Intelligent Perception and Cognition, \\University of Science and Technology of China, Hefei, China}
\address[B]{Anhui Province Key Laboratory of Biomedical Imaging and Intelligent Processing, \\Institute of Artificial Intelligence, Hefei Comprehensive National Science Center, Hefei, China}
\address[C]{Tianjin University, Tianjin, China}
\address[D]{Beijing Tiantan Hospital, Capital Medical University, Beijing, China}


\begin{abstract}
   Multi-modal Magnetic Resonance Imaging (MRI) is imperative for accurate brain tumor segmentation, offering indispensable complementary information. Nonetheless, the absence of modalities poses significant challenges in achieving precise segmentation. Recognizing the shared anatomical structures between mono-modal and multi-modal representations, it is noteworthy that mono-modal images typically exhibit limited features in specific regions and tissues. In response to this, we present Anatomical Consistency Distillation and Inconsistency Synthesis (ACDIS), a novel framework designed to transfer anatomical structures from multi-modal to mono-modal representations and synthesize modality-specific features.
    ACDIS consists of two main components: Anatomical Consistency Distillation (ACD) and Modality Feature Synthesis Block (MFSB).
    ACD incorporates the Anatomical Feature Enhancement Block (AFEB), meticulously mining anatomical information. Simultaneously, Anatomical Consistency ConsTraints (ACCT) are employed to facilitate the consistent knowledge transfer, i.e., the richness of information and the similarity in anatomical structure, ensuring precise alignment of structural features across mono-modality and multi-modality. 
    Complementarily, MFSB produces modality-specific features to rectify anatomical inconsistencies, thereby compensating for missing information in the segmented features. Through validation on the BraTS2018 and BraTS2020 datasets, ACDIS substantiates its efficacy in the segmentation of brain tumors with missing MRI modalities.
\end{abstract}

\end{frontmatter}


\section{Introduction}
Brain tumors have a profound impact on overall health, underscoring the critical necessity for precise segmentation crucial in both diagnosis and the monitoring of treatment outcomes. The advanced segmentation of brain tumors often relies on the integration of various Magnetic Resonance Imaging (MRI) sequences, encompassing FLuid Attenuation Inversion Recovery (FLAIR), contrast-enhanced T1-weighted (T1ce), T1-weighted (T1), and T2-weighted (T2) modalities. These sequences collectively provide complementary information essential for achieving precise diagnostic outcomes. However, the clinical reality presents challenges, as the simultaneous availability of all these modalities cannot be guaranteed. This uncertainty gives rise to difficulties in achieving accurate tumor segmentation. For example, the unavailability of the T1c modality may be attributed to patient allergies to contrast agents, while other modalities may be inaccessible due to disparities in MRI parameters. Consequently, addressing the segmentation of brain tumors in the absence of certain modalities has emerged as a critical imperative in clinical settings~\cite{azad2022medical}.

\begin{figure}[t]
    \includegraphics[width=\linewidth]{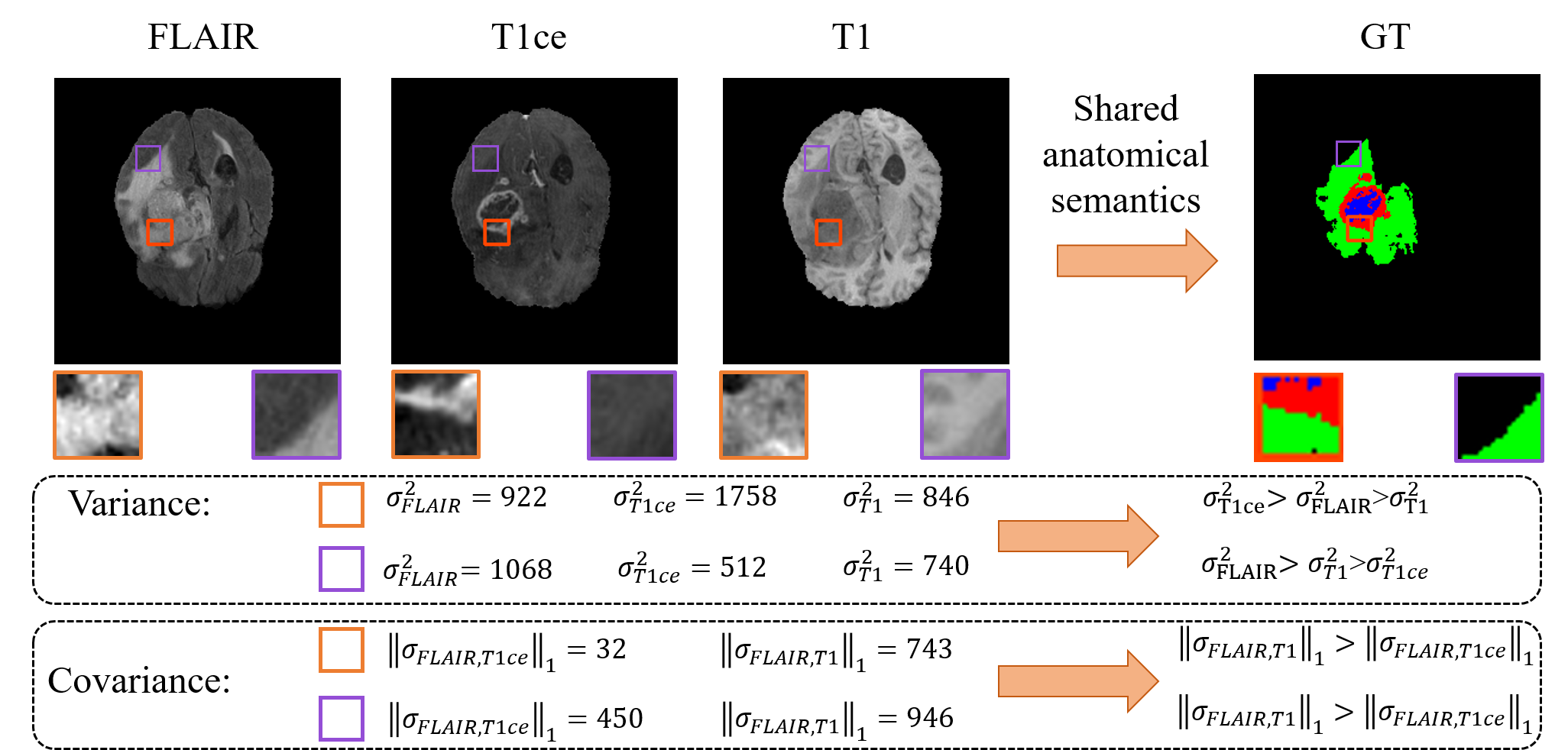}
    \vspace{0.2cm}
    \caption{Different modalities share the same anatomical semantics, but they vary in terms of pixel intensity visualization. Typically, 1. The modality with richer information and clearer anatomical structure exhibits greater variance. 2. Greater anatomical similarity between modalities results in higher absolute covariance values.}
    \vspace{0.6cm}
    \label{fig:moti}
\end{figure}

Current methods for handling missing modalities primarily focus on two strategies: modality synthesis~\cite{lee2020assessing,zhou2020hi,yu2022mousegan++,yang2023learning,kim2024adaptive} and common latent space modeling~\cite{ding2021rfnet,zhou2021latent,zhang2022mmformer,qiu2023scratch,wei2023mmanet}. Modality synthesis methods utilize generative adversarial networks or diffusion models to generate missing modalities. 
Nevertheless, these generated modalities lack comprehensive modality-specific biological information~\cite{lee2020assessing}, limiting their effectiveness in achieving accurate segmentation.
On the other hand, common latent space modeling aims to project each modality into a shared latent space, where modality features are fused to obtain the segmentation results. 
This strategy incorporates knowledge distillation to guide the extraction and aggregation of mono-modal and multi-modal features~\cite{wang2021acn,qiu2023scratch,wei2023mmanet}. Despite these advancements, current methods tend to uniformly transfer knowledge between mono-modality and multi-modality, neglecting the distinct characteristics that are unique to each modality. Crucially, while these modalities share consistent anatomical semantics like tumor or tissue types~\cite{yang2023directional,akeret2023anatomy}, they exhibit significant inconsistencies in pixel intensity visualization. 

To address these issues, we propose the Anatomical Consistency Distillation and Inconsistency Synthesis (ACDIS) through elaborately combined knowledge distillation and feature synthesis for brain tumor segmentation with missing modalities. Our method comprises two core components: Anatomical Consistency Distillation (ACD) and Modality Feature Synthesis Block (MFSB). 
ACD focuses on transferring consistent anatomic insights from multi-modality to mono-modality, enhancing the representation of anatomical structures. 
It utilizes Anatomical Consistency ConsTraints (ACCT), emphasizing the transmission of knowledge based on variance and covariance within a local window, which highlights information richness and anatomical structural similarity. These constraints ensure that the transfer of knowledge is not only rich in detail but also precisely aligned with the structural similarities inherent in anatomical features. The Anatomical Feature Enhancement Block (AFEB) complements this by deeply extracting mono-modal anatomical information, cooperating with ACCT to refine and enrich the mono-modal representations.

On the other hand, MFSB synthesizes modality-specific features that address the inconsistencies often seen in anatomical information due to varying pixel intensities. It generates a modality-specific style for transferring existing mono-modal features to simulate the missing modalities, compensating for the discrepancies in anatomical structure visualization.

Our main contributions can be summarized as follows:
\begin{itemize}
    \item We investigate the anatomical consistency and inconsistency between the mono-modality and the multi-modality. Based on them, we propose ACDIS that leverages the consistent anatomical insights and synthesizes the inconsistent anatomical features.
    \item ACDIS incorporates ACD with a focus on variance and covariance in local windows, conveying rich information and emphasizing anatomical structural similarity, thereby enhancing mono-modal representation. MFSB is employed to learn modality-specific styles, addressing the gap in missing information.
    \item ACDIS consistently outperforms state-of-the-art methods for brain tumor segmentation with missing modalities on BraTS2018 and BraTS2020.
\end{itemize}


\section{Related Work}
\subsection{Brain Tumor Segmentation}
Convolutional Neural Networks (CNNs) and Transformers are both potent architectures for feature extraction in brain tumor segmentation tasks. Within the domain of CNNs, several methods build upon the foundational U-Net model~\cite{ronneberger2015u}, focusing on enhancements such as 2D/3D operations~\cite{cciccek20163d}, cascaded networks~\cite{jiang2020two, zhou2020one}, skip connections~\cite{zhou2019unet++} and automated adaptations~\cite{isensee2021nnu}. Transformer architectures, known for their ability to model long-range dependencies, offer significant advantages for capturing the global context necessary for accurate segmentation. Chen \emph{et al.}~\cite{chen2021transunet} incorporated Transformer layers on top of CNN features to utilize global contextual information. Inspired by the hierarchical Swin Transformer~\cite{liu2021swin}, Cao \emph{et al.}~\cite{cao2023swin} developed a U-shaped architecture using Swin Transformer blocks, which facilitates robust feature extraction by leveraging the advantages of both CNNs and Transformers. This integration of CNN and Transformer is followed by numerous methods~\cite{zhang2021transfuse, wang2021transbts, li2022transbtsv2, hatamizadeh2022unetr, hatamizadeh2022swin}. However, the phenomenon of missing modalities prevents them from extracting complete complementary information, posing an obstacle for accurate brain tumor segmentation.

\subsection{Multi-modal Learning for Missing Modalities}
Recently, many multi-modal learning methods adopt modality synthesis and common latent space modeling to overcome the missing modality issue. Modality synthesis~\cite{lee2020assessing,yang2023learning} methods that generate missing modalities and train a segmentation model with complete modalities. HiNet and MouseGAN++ employ multiple specific generative adversarial networks to separately generate missing modalities~\cite{zhou2020hi,yu2022mousegan++}.
M3AE reconstructs multi-modal substitutes within multi-modal mask image modeling strategy~\cite{liu2023m3ae}.
Kim \textit{et al.} propose 3D latent diffusion models combined with conditioning using the target modality allows generating high-quality target modality in 3D~\cite{kim2024adaptive,rombach2022high}. These synthesis methods often require training their networks on various combinations of missing modalities, which can be computationally intensive. Moreover, the segmentation performance is closely tied to the quality of the generated modalities, which may often contain unexpected noise.

Common latent space modeling methods~\cite{ding2021rfnet,konwer2023enhancing,wang2023multi,zhang2024tmformer} that project all available modalities into a shared latent space, and then fuse all modalities to produce a segmentation map.
RFNet, LCRL, and mmFormer employ mono-modal segmentation or reconstruction tasks to improve the mono-modal representation in the latent space~\cite{ding2021rfnet,zhou2021latent,zhang2022mmformer}.
ShapSpec further enhances modality features through feature domain classification and feature distribution alignment~\cite{wang2023multi}.
Some methods, such as~\cite{chen2019robust,yang2022d}, disentangle modalities into several attributes and process them separately. 
Konwer \textit{et al.} adopt meta-learning and adversarial learning strategies to enhance modality-agnostic representations~\cite{konwer2023enhancing}. TMFormer~\cite{zhang2024tmformer} employs token merging strategy to obtain compact representation in mono-modal and multi-modal token sequences.
While these methods primarily focus on improving representations in scenarios involving missing modalities, they overlook the intrinsic biological information mined in multi-modal modalities, which can boost the mono-modal representations by modeling data distribution relationships between mono-modality and multi-modality.

\begin{figure*}[t]
    \centering
    \includegraphics[width=\textwidth]{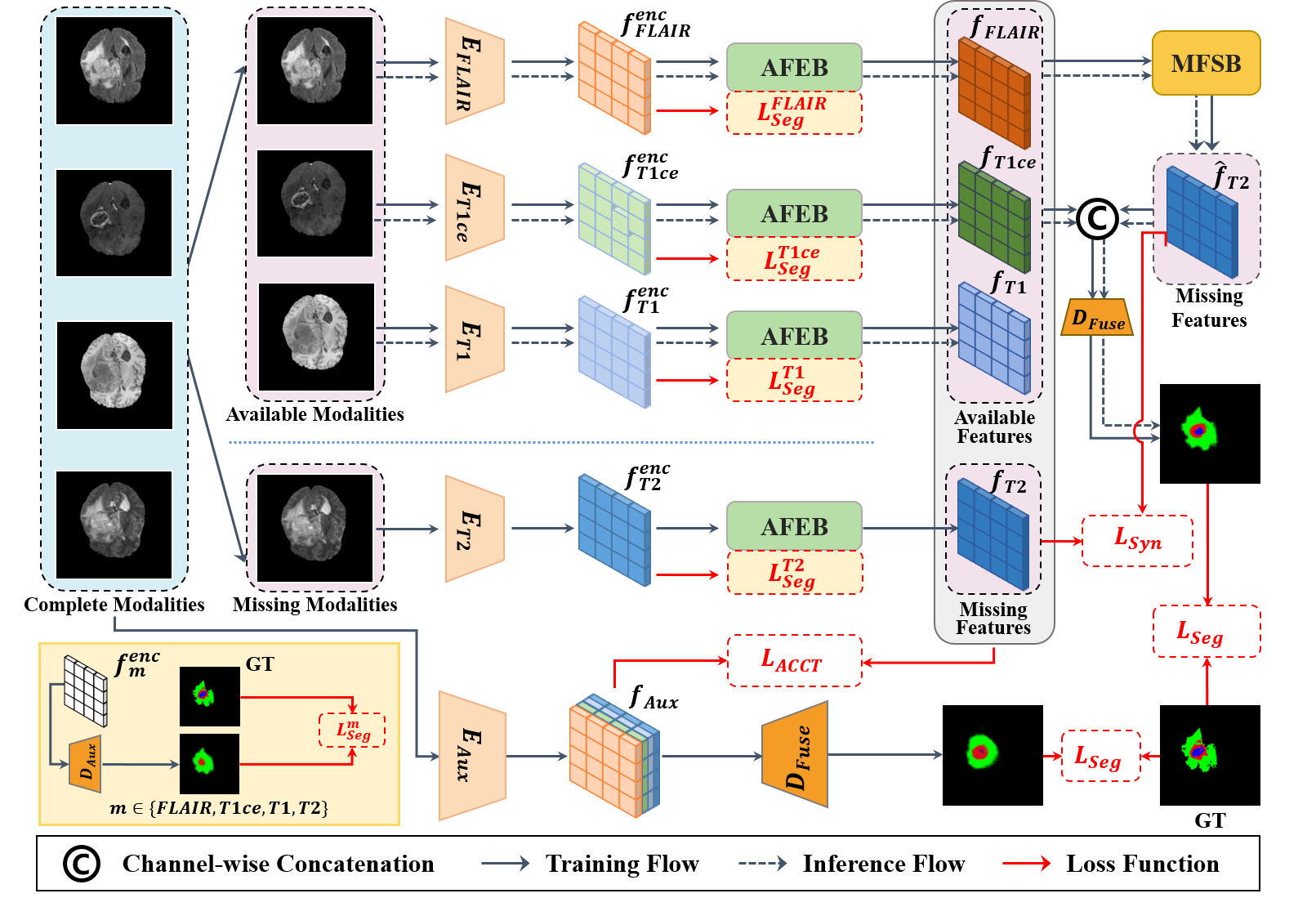}
    \vspace{0.2cm}
    \caption{Overall architecture of our ACDIS. During the training phase, ACDIS comprises four mono-encoders for extracting mono-modal features, one auxiliary encoder dedicated to consistency distillation in conjunction with the proposed AFEB and ACCT, one auxiliary decoder for obtaining individual mono-modal segmentation (denoted in the yellow box), one MFSB designed to synthesize features for missing modalities, and one fusion decoder responsible for generating the final segmentation result. The auxiliary encoder and decoder are discarded during the inference phase.}
    \vspace{0.6cm}
    \label{fig:framework}
\end{figure*}

\subsection{Knowledge Distillation}
Cross-modal distillation has been a long-standing approach for transferring specific knowledge between different modalities~\cite{gupta2016cross}. KDNet~\cite{hu2020knowledge} and ACN~\cite{wang2021acn} transfer feature distributions from the multi-modal network to the mono-modal network, focusing on latent features or soft segmentation masks, respectively.  
MMANet~\cite{wei2023mmanet} focuses on distilling information from hard samples that lie near the decision boundary. Additionally, GSS~\cite{qiu2023scratch} uses an ensemble of mono-modal soft segmentation masks as the teacher's output for distillation. However, these methods obscure the distinction between what should be consistent and what should be inconsistent across modalities. It is crucial to maintain consistency in anatomical structures across mono-modal and multi-modal data while accepting inconsistency in pixel intensities. Introducing conflicting information that contradicts specific imaging technologies adversely affects the student model’s performance.


\section{Method}
Our framework is illustrated in Figure~\ref{fig:framework}. Each modality is sent into individual encoder $\{E_m\}_{m\in\{FLAIR, T1ce, T1, T2\}}$ for extracting hierarchical features, which are then fed into the auxiliary decoder $D_{Aux}$ to yield the mono-modal segmentation. To enhance the representation of mono-modal anatomical structures, the mono-modal hierarchical features are passed through AFEB to elaborately model spatial structural dependencies in collaboration with ACCT. ACCT introduces an auxiliary encoder $E_{Aux}$ that provides multi-modal features with comprehensive anatomical information. ACCT and AFEB are essential components of our ACT for ensuring anatomical consistency. After being processed by ACD, the individual features contain richer anatomical structure information. The MFSB utilizes available mono-modal features to synthesize the missing ones, thus addressing anatomical inconsistencies. The completed multi-modal features are then fed into the decoder $D_{Fuse}$ to obtain the final segmentation.

\subsection{Anatomical Consistency Distillation (ACD)}
The ACD consists of ACCT and AFEB, where AFEB refines the mono-modal representation cooperating with ACCT.

\subsubsection{Anatomical Consistency ConsTraints (ACCT)}
While different modalities may exhibit distinct pixel intensities attributable to their respective imaging technologies, they share consistent anatomical semantics in terms of relative positions, especially when registered to standard atlases. Multi-modal data inherently encapsulates a more comprehensive anatomical information compared to mono-modal data. Consequently, our approach focuses on distilling anatomical structural information from the multi-modal data into the mono-modal data. 

As shown in Figure~\ref{fig:moti}, windows with higher variances $\sigma^2$ typically contain richer information related to the segmentation mask, such as more pronounced contrast at the segmentation boundaries. This observation leads us to utilize variance as a key feature in our distillation process. This is illustrated as
\begin{align}
    L_{Var} = 1-\frac{{2\sigma_{f_m}\sigma_{f_{Aux}} + \epsilon}}{{\sigma_{f_m}^2 + \sigma_{f_{Aux}}^2 + \epsilon}},
    \label{eq:var}
\end{align}%
where $f_m$ and $f_{Aux}$ represent the mono-modal features and multi-modal features, respectively. The multi-modal features $f_{Aux}$ are fixed to serve as the label for the mono-modal features $f_m$. The term $\epsilon=1e$-$6$  is included to prevent division by zero.



The constraint on the variance pushes the mono-modal representation to learn information richness of the multi-modal representation. Secondly, we employ the covariance to push the mono-modal representation learn the anatomical structure of the multi-modal representation, which quantifies the variation trends among pixel values. When two local windows exhibit similar gradients in corresponding positions, the covariance tends to increase. The covariance of two windows can be described as
\begin{align}
    \sigma_{f_mf_{Aux}} = \frac{1}{n-1} \sum_{i=1}^{n} (f_{m_i} - \bar{f}_m)(f_{{Aux}_i} - \bar{f}_{Aux}).
\end{align}

Since FLAIR and T1ce MRI modalities often exhibit opposite pixel intensities due to their underlying imaging principles and the use of contrast agents, this may cause the feature pixel intensity demonstrating the opposite variational trend. Therefore, we use $||\cdot||_1$ in the constraint to only focus on the gradient magnitude, which is presented as
\begin{align}
    L_{Covar} = 1-\frac{{||\sigma_{win_1 win_2}||_1 + \epsilon}}{{\sigma_{win_1}\sigma_{win_2} + \epsilon}}.
    \label{eq:covar}
\end{align}

Note that we use Sigmoid function to normalize the mono-modal features $f_m$ and the multi-modal features $f_{Aux}$ before sending into Equations \ref{eq:var} and \ref{eq:covar}. The loss in ACCT can be presented as
\begin{align}
    L_{ACCT} = L_{Var} + L_{Covar}.
    \label{eq:ACCT}
\end{align}

\subsubsection{Anatomical Feature Enhancement Block (AFEB)}
\begin{figure}[t]
    \centering
    \includegraphics[width=\linewidth]{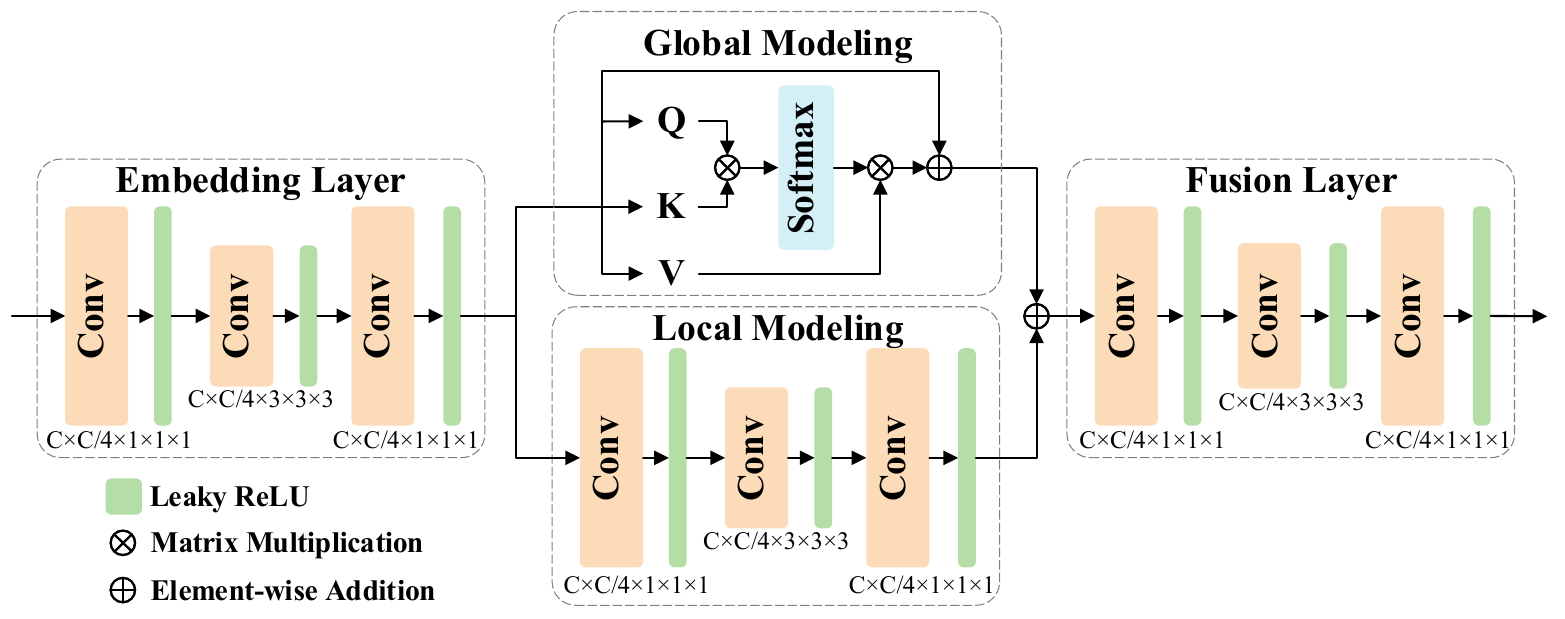}
    \vspace{0.2cm}
    \caption{Anatomical Feature Enhancement Block (AFEB).}
    \vspace{0.6cm}
    \label{fig:AFEB}
\end{figure}
In order to effectively assimilate anatomical structural knowledge from multi-modal data, we devise an Anatomical Feature Enhancement Block (AFEB) to facilitate the extraction of spatial anatomical information within the mono-modal domain and establish data relationships between mono-modal and multi-modal representations. 
As depicted in Figure~\ref{fig:AFEB}, we firstly employ embedding layers to project the mono-modal features into a multi-modal space, facilitating the modeling of relationships between mono-modal and multi-modal representations. Subsequently, we employ both an attention layer and a convolutional layer to systematically explore spatial dependencies in both global and local aspects. Finally, we fuse the global and the local features, obtaining the enhanced mono-modal feature $f_m$, where $m\in\{FLAIR, T1ce, T1, T2\}$.

The enhanced mono-modal feature $f_m$ is further constrained with proposed ACCT loss $L_{ACCT}$, which distills the multi-modality features to $f_m$. In this way, the ACD framework enhances the mono-modal representation by leveraging the underlying anatomical structure consistency.



\subsection{Modality Feature Synthesis Block (MFSB)}
\begin{figure}[t]
    \centering
    \includegraphics[width=\linewidth]{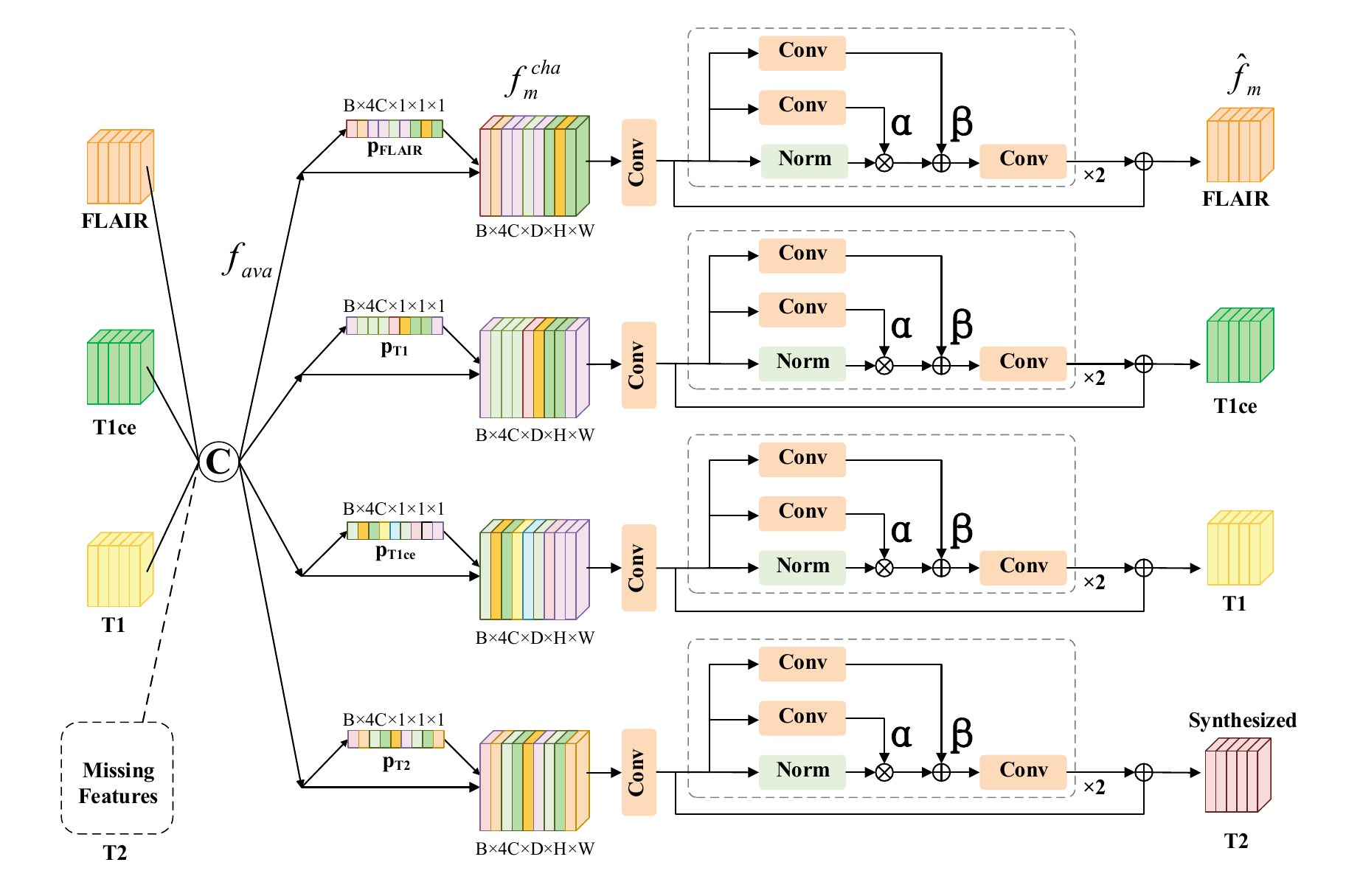}
    \vspace{0.2cm}
    \caption{Modality Feature Synthesis Block (MFSB).}
    \vspace{0.6cm}
    \label{fig:MFSB}
\end{figure}
Building upon the foundation laid by ACD, we introduce the Multi-Modal Feature Synthesis Block (MFSB) to address the anatomical inconsistencies, specifically focusing on the feature intensities of missing modalities. As illustrated in Figure.~\ref{fig:MFSB}, all available mono-modal features are concatenated in channel dimension, denoted as $f_{ava}=\text{Concat}(f_{FLAIR}, f_{T1ce}, f_{T1}, f_{T2})\in\mathcal{R}^{B\times4C\times D \times H \times W} $. In cases where the corresponding modality $f_m$ is absent, we compensate it by assigning zero values and proceed to synthesize it in the subsequent steps.

\begin{figure*}[htb]
    \centering
    \includegraphics[width=0.95\textwidth]{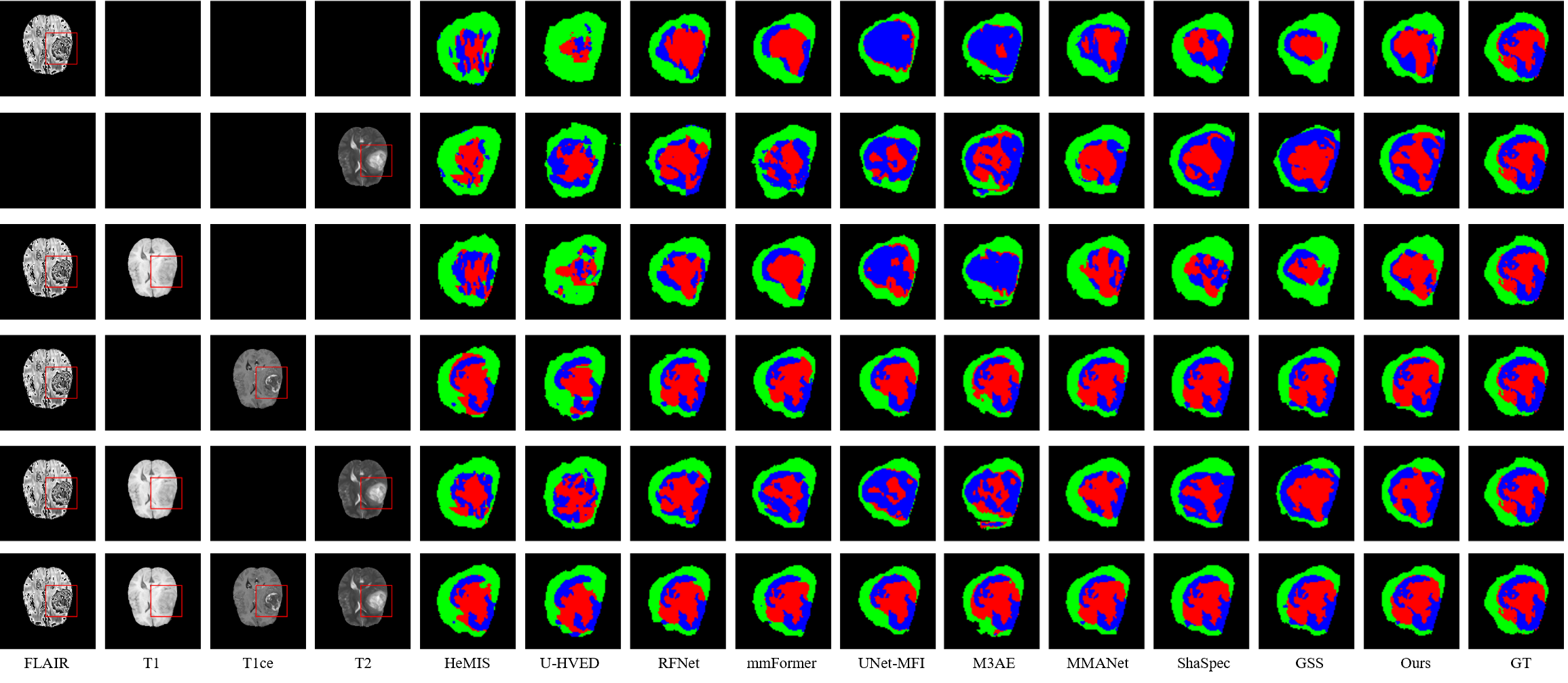}
    \vspace{0.2cm}
    \caption{Segmentation results of different methods with various available modalities on BraTS2020.}
    \vspace{0.6cm}
    \label{fig:main}
\end{figure*}

For each modality synthesis $\hat{f}_m$, we learn the prior weight $p_m$, which serves to model inter-modal significance along the channel dimension, thereby amplifying the most relevant mono-modal features. A dimension-reducing convolution operation $W_2 \in \mathcal{R}^{C\times4C}$ is employed to squeeze the amplified features. The processes can be expressed as
\begin{align}
    p_m &= \delta (W_1 \text{GAP}(f_{ava})), \\
    f_{m}^{cha} &= W_2 p_m f_{ava},
    \label{eq:se}
\end{align}
where $\delta$ is the sigmoid function, $W_1 \in \mathcal{R}^{4C\times4C}$, $\text{GAP}(\cdot)$ represents the global average pooling function, and $f_{m}^{cha}\in \mathcal{R}^{B\times C\times D \times H \times W}$. Subsequently, we learn the modal-specific style on the spatial dimension. The process is expressed as
\begin{align}
    \alpha_m &= \delta (W_3 f_{m}^{cha}), \\
    \beta_m &= \delta (W_4 f_{m}^{cha}), \\
    \hat{f}_{m} &= (1+\alpha_m) f_{m}^{cha} + \beta_m,
    \label{eq:style}
\end{align}
where $\alpha_m \in \mathcal{R}^{B\times C\times D \times H \times W}$, $\beta_m \in \mathcal{R}^{B\times C\times D \times H \times W}$ represent learned affine transformations that encapsulate the modality-specific styles, and $W_3 \in \mathcal{R}^{C\times C}$ and $W_4 \in \mathcal{R}^{C\times C}$ denote corresponding convolutional operations, respectively. 
In this context, $\alpha_m$ serves to adjust the magnitude of synthesized features for the missing modalities, while $\beta_m$ facilitates the shifting of these features to a suitable space. 
In the training stage, we can obtain the complete modality-specific features with complete modalities input. We use the Mean Square Error function to supervise the synthesis of $\hat{f}_{m}$, prioritizing pixel-wise intensity, aligning with the objective of inconsistency synthesis. The loss is presented as
\begin{align}
    L_{Syn} = || \hat{f}_{m}-f_{m} ||_2.
    \label{eq:mse}
\end{align}

We do not employ the adversarial loss~\cite{goodfellow2020generative}, since the primary objective of the discriminator is to enhance synthesis diversity, while our goal is to generate the missing features with high fidelity. In this manner, we synthesize the feature intensities of missing modalities that compensate the anatomical inconsistent information. 

Since we obtain the complete multi-modal features $f_{comp} \in\mathcal{R}^{B\times4C\times D \times H \times W}$ using our MFSB, we forward $f_{comp}$ to the decoder $D_{Fuse}$ to derive the final segmentation results. The implementation of this decoder is similar to the 3D U-Net architecture.

\subsection{The Overall Loss}
In line with previous works such as \cite{ding2021rfnet,zhang2022mmformer}, we employ the weighted cross-entropy loss denoted as $L_{WCE}$ and the Dice loss denoted as $L_{Dice}$ to align our predictions with the corresponding ground-truth segmentation maps. These losses are formulated as
\begin{align}
    L_{Seg} = \sum_{i=1}^N (L_{WCE}(y_i, \hat{y}_i) + L_{Dice}(y_i, \hat{y}_i)),
    \label{eq:seg}
\end{align}
where $\hat{y}_i$ and $y_i$ are the predicted segmentation and the corresponding ground-truth, respectively. The parameter $N=6$ signifies that we are predicting six segmentation maps. These include four segmentation maps derived from a shared auxiliary decoder $D{Aux}$, each taking one of the four modalities as input, one segmentation map obtained from a fusion decoder $D_{Fuse}$ that utilizes multi-modal features from the multi-modal encoder $E_{Aux}$, and one segmentation map generated by the shared fusion decoder $D_{Fuse}$ using compensated features from our MFSB.

Consequently, by combining Equations \ref{eq:ACCT}, \ref{eq:mse}, and \ref{eq:seg}, the overall loss is expressed as
\begin{align}
    L_{Overall} = L_{ACCT} + L_{Seg} + L_{Syn}.
    \label{eq:over}
\end{align}

\begin{table*}[h]
    \centering
    \resizebox{\textwidth}{!}{%
    \begin{tabular}{c|c|c|c|c|c|c|c|c|c|c|c|c|c|c|c|c|c}
        \hline
        \multirow{4}{*}{M}  & FLAIR   & $\bullet$ & $\circ$   & $\circ$   & $\circ$   & $\bullet$ & $\bullet$ & $\bullet$ & $\circ$   & $\circ$   & $\circ$   & $\bullet$ & $\bullet$ & $\bullet$ & $\circ$   & $\bullet$ & \multirow{4}{*}{AVG}   \\
                            & T1ce & $\circ$   & $\bullet$ & $\circ$   & $\circ$   & $\bullet$ & $\circ$   & $\circ$   & $\bullet$ & $\bullet$ & $\circ$   & $\bullet$ & $\bullet$ & $\circ$   & $\bullet$ & $\bullet$ &       \\
                            & T1  & $\circ$   & $\circ$   & $\bullet$ & $\circ$   & $\circ$   & $\bullet$ & $\circ$   & $\bullet$ & $\circ$   & $\bullet$ & $\bullet$ & $\circ$   & $\bullet$ & $\bullet$ & $\bullet$ &       \\
                            & T2  & $\circ$   & $\circ$   & $\circ$   & $\bullet$ & $\circ$   & $\circ$   & $\bullet$ & $\circ$   & $\bullet$ & $\bullet$ & $\circ$   & $\bullet$ & $\bullet$ & $\bullet$ & $\bullet$ &       \\ \hline
        \multirow{10}{*}{WT} & HeMiS     & 71.60             & 67.71             & 68.96             & 68.19             & 69.17             & 68.67             & 69.83             & 69.01             &                          69.78             & 69.40             & 70.21            & 71.28              & 70.73             & 71.58             & 72.06             & 69.88  \\
                            & U-HVED    & 69.85             & 46.82             & 46.77             & 54.03             & 61.45             & 58.25             & 64.50             & 62.91             & 65.76             & 64.29             & 66.99             & 69.70             & 68.38             & 70.35             & 71.41             & 62.76        \\
                            & RFNet     & 86.42             & \textbf{77.34} & \underline{76.46}             & 86.21 & \underline{89.55} & 89.30             & 89.35 & \underline{81.00} & 87.45 & 87.95    & 90.39    & 90.20 & 90.42 & 88.59 & 90.77 & 86.76        \\
                            & UNet-MFI  & 82.27             & 73.18             & 72.10             & 82.45             & 83.64             & 84.34             & 84.85             & 77.30             & 83.44             & 83.52             & 85.45             & 85.70             & 85.85             & 84.20             & 84.93             & 82.21        \\ 
                            & mmFormer  & 82.40             & 74.25             & 74.37             & 83.07             & 84.54             & 84.61             & 85.82             & 77.98             & 84.05             & 84.00             & 85.34             & 86.11             & 86.22             & 84.64             & 86.38             & 82.92        \\ 
                            & MMANet      & 83.21 & 64.46             & 64.35 & 82.88             & 86.74             & 87.64 & 87.47             & 70.98             & 85.64             & 85.33             & 87.92             & 89.37             & 87.97             & 86.38    & 89.25             & 82.64        \\ 
                            & ShaSpec      & 84.10 & 70.36             & 70.23 & 83.95             & 88.65             & 89.07 & 88.70             & 76.02             & 86.21             & 87.15             & 89.47             & 89.43             & 90.07             & 87.62    & 90.11             & 84.74        \\
                            & M3AE      & \underline{86.53} & 73.85             & \textbf{76.71} & 86.09             & 89.48             & 89.38 & 89.25             & 78.11             & 87.37             & 87.20             & 89.99             & 90.18             & 90.42             & 88.61    & 90.56             & 86.25        \\
                            & GSS*      & 86.36 & 72.45             & 73.33 & \underline{86.76}             & 89.51             & \underline{90.65} & \underline{90.91}             & 78.38             & \underline{88.87}             & \textbf{90.05}             & \underline{91.14}             & \textbf{93.25}             & \underline{90.67}             & \underline{89.43}    & \textbf{93.28}             & \underline{87.00}        \\
                            & Ours      & \textbf{88.74}    & \underline{76.59}    & 76.36    & \textbf{87.41}    & \textbf{91.14}    & \textbf{91.71}    & \textbf{91.49}    & \textbf{81.66}    & \textbf{89.83}    & \underline{89.61} & \textbf{92.16} & \underline{92.40}    & \textbf{92.14}    & \textbf{90.62}             & \underline{92.82}    & \textbf{88.31}        \\ \hline
        \multirow{10}{*}{TC} & HeMiS     & 53.43 & 51.41 & 51.56 & 51.11 & 51.70 & 51.08 & 51.85 & 51.88 & 52.35 & 51.51 & 52.95 & 53.76 & 52.97 & 54.38 & 55.03 & 52.46   \\
                            & U-HVED    & 34.62 & 35.51 & 27.30 & 37.67 & 42.15 & 38.26 & 43.41 & 44.93 & 47.53 & 44.97 & 49.13 & 51.30 & 49.40 & 52.72 & 54.17 & 43.53         \\
                            & RFNet     & 65.04 & \underline{82.37} & \underline{64.31} & 68.47 & 84.69 & 71.45 & 72.62 & \underline{83.15} & 84.06 & 72.11 & 84.71 & 84.70 & 74.28 & 84.11 & 84.74 & 77.39      \\
                            & UNet-MFI  & 63.94 & 77.63 & 59.38 & 68.05 & 79.92 & 68.23 & 70.72 & 77.61 & 80.09 & 70.21 & 80.03 & 80.94 & 71.40 & 80.75 & 81.28 & 74.01      \\
                            & mmFormer  & 66.19 & 77.96 & 61.17 & 69.18 & 80.36 & 69.58 & 71.55 & 79.93 & 80.79 & 70.90 & 80.18 & 81.31 & 72.02 & 81.12 & 81.22 & 74.90  \\
                            & MMANet     & 65.79 & 74.05 & 57.47 & 70.21 & \underline{85.21} & 72.73 & 72.65 & 80.99 & 85.24 & 71.88 & 85.87 & 85.68 & 74.48 & \textbf{86.39} & 86.63 & 76.99   \\
                            & ShaSpec    & 66.06 & 77.34 & 59.57 & 68.29 & 85.02 & 72.87 & 72.95 & 80.84 & 84.46 & 72.28 & 85.83 & 85.57 & 74.74 & 85.47 & 86.50 & 77.19         \\
                            & M3AE      & \underline{68.04} & 81.39 & \textbf{66.00} & \underline{70.27} & 82.01 & \underline{73.82} & \underline{74.95} & 82.39 & 83.01 & 72.54 & 82.44 & 83.06 & \underline{75.09} & 84.06 & 84.40 & 77.56  \\
                            & GSS*    & 67.33 & 78.43 & 61.67 & 70.10 & 84.88 & 73.45 & 74.17 & 82.20 & \underline{86.12} & \textbf{74.18} & \underline{86.50} & \textbf{88.39} & 74.33 & 86.28 & \textbf{88.67} & \underline{78.45}         \\
                            & Ours      & \textbf{71.05} & \textbf{82.61} & 64.08 & \textbf{72.03} & \textbf{85.84} & \textbf{75.31} & \textbf{75.00} & \textbf{84.34} & \textbf{86.34} & \underline{73.57} & \textbf{86.95} & \underline{86.62} & \textbf{76.43} & \underline{86.35} & \underline{86.92} & \textbf{79.56}    \\ \hline
        \multirow{10}{*}{ET} & HeMiS     & \textbf{43.77} & 42.41 & \textbf{41.59} & 41.45 & 41.83 & 40.29 & 41.19 & 42.08 & 42.39 & 41.00 & 43.67 & 44.16 & 42.95 & 45.27 & 46.33 & 42.69                 \\
                            & U-HVED    & 12.88 & 24.94 &  7.27 & 24.26 & 30.02 & 21.95 & 29.40 & 33.64 & 36.18 & 32.12 & 39.39 & 40.91 & 38.09 & 43.18 & 45.33 & 30.64       \\
                            & RFNet     & 40.47 & \textbf{74.27} & 37.51 & 43.59 & \textbf{76.45} & 43.81 & 46.99 & 75.22 & 73.94 & 46.37 & \underline{77.01} & 76.38 & 48.95 & 76.38 & 76.64 & 60.93      \\
                            & UNet-MFI  & 39.70 & 69.42 & 29.38 & \textbf{46.00} & 70.13 & 40.06 & 48.69 & 69.25 & 72.32 & 45.71 & 71.28 & 70.88 & 46.55 & 72.00 & 71.41 & 57.52      \\
                            & mmFormer  & 40.47 & 68.91 & 33.97 & 45.61 & 69.81 & 43.63 & 48.09 & 71.10 & 70.72 & 45.92 & 70.08 & 71.60 & 48.38 & 70.65 & 71.36 & 58.02   \\
                            & MMANet     & 36.40 & 66.07 & 28.22 & 41.91 & 70.43 & 41.50 & 44.23 & 68.87 & 71.86 & 43.58 & 71.94 & 71.98 & 44.60 & 71.81 & 72.43 & 56.39                 \\
                            & ShaSpec    & 38.92 & 66.93 &  31.75 & 42.58 & 70.85 & 43.22 & 45.62 & 69.92 & 71.04 & 44.21 & 72.08 & 71.09 & 46.22 & 71.81 & 72.02 & 57.22       \\
                            & M3AE      & 40.49 & 72.43 & \underline{39.93} & 45.97 & 74.66 & 43.20 & 47.30 & \underline{75.42} & \underline{76.81} & 46.63 & 75.94 & \textbf{77.08} & 48.19 & \underline{77.40} & \underline{78.00} & \underline{61.30}   \\
                            & GSS*    & \underline{42.03} & 69.46 &  35.30 & 45.60 & 74.20 & \underline{47.78} & \underline{49.33} & 74.41 & 74.78 & \textbf{48.94} & 76.12 & 76.32 & \underline{50.02} & 76.98 & 75.95 & 61.15       \\
                            & Ours      & 41.50 & \underline{73.42} & 36.88 & \underline{45.99} & \underline{76.35} & \textbf{47.87} & \textbf{50.09} & \textbf{77.24} & \textbf{76.83} & \underline{48.84} & \textbf{79.55} & \underline{77.07} & \textbf{52.75} & \textbf{78.36} & \textbf{79.05} & \textbf{62.79}   \\ \hline
        \end{tabular}%
        }
    \vspace{0.2cm}
    \caption{Performance comparison (DSC\%) with SOTA methods, including HeMiS, U-HVED, RFNet, UNet-MFI, mmFormer, MMANet, ShaSpec, M3AE, and GSS on BraTS2020. Available and missing modalities are represented by $\bullet$ and $\circ$, respectively. The `*' demonstrates that its codes and results are reproduced by ourselves.}
    \vspace{0.6cm}
    \label{tab:main_table}
    \end{table*}

\begin{table}[t]
    \centering
    \resizebox{\linewidth}{!}{
    \begin{tabular}{@{}c|cccc|ccc@{}}
    \hline
    \multicolumn{5}{c|}{Consistency Distillation} & \multicolumn{3}{c}{Dice} \\ \hline
    \multicolumn{1}{c|}{\multirow{2}{*}{w/o AFEB}} & \multicolumn{4}{c|}{w/ AFEB} & \multirow{2}{*}{WT} & \multirow{2}{*}{TC} & \multirow{2}{*}{ET} \\ \cline{2-5}
    \multicolumn{1}{c|}{} & $L_{MSE}$  & $L_{KL}$ & $L_{Var}$ & $L_{Covar}$ & & & \\ \hline
    \checkmark &        &    &     &       &  83.34      &   75.36     &  56.22      \\
    &    \checkmark     &    &     &       &  84.85      &   76.68     &  57.77      \\
         &    & \checkmark   &     &       &  74.35      &  62.86      &  47.29      \\
         &    &    &  \checkmark   &       &  85.97      &   77.23     &   59.71     \\
         &    &    &     & \checkmark      &  86.04      &  77.98      &   60.23     \\
         &    &    &  \checkmark   &  \checkmark     &   \textbf{86.61}     &   \textbf{78.56}     &   \textbf{61.09}     \\ \hline
    \end{tabular}}
    \vspace{0.2cm}
    \caption{Ablation study on consistency distillation.}
    \vspace{0.6cm}
    \label{tab:kd}
\end{table}

\begin{table}[t]
    \centering
    \resizebox{\linewidth}{!}{
    \begin{tabular}{@{}c|cccc|ccc@{}}
    \hline
    \multicolumn{5}{c|}{Inconsistency Synthesis} & \multicolumn{3}{c}{Dice} \\ \hline
    \multicolumn{1}{c|}{\multirow{2}{*}{w/o MFSB}} & \multicolumn{4}{c|}{w/ MFSB} & \multirow{2}{*}{WT} & \multirow{2}{*}{TC} & \multirow{2}{*}{ET} \\ \cline{2-5}
    \multicolumn{1}{c|}{} & $L_{Adver}$  & $L_{Var}$ & $L_{Covar}$ & $L_{Syn}$ & & & \\ \hline
    \checkmark &        &    &     &       &  86.61      &   78.56     &  61.09      \\
    & \checkmark     &    &     &       &  86.89      &  78.73      &   61.41     \\
    &    & \checkmark   & \checkmark  &        & 87.16      &  78.67      &  61.55   \\
    &    &     &    &  \checkmark      & \textbf{88.31}      &  \textbf{79.56}      &  \textbf{62.79}   \\
    &    \checkmark     &    &     &  \checkmark     &  88.00      &  79.21      &  62.34      \\
    &    & \checkmark   & \checkmark    & \checkmark      &  88.11      &  79.42      &  62.50      \\ \hline
    \end{tabular}}
    \vspace{0.2cm}
    \caption{Ablation study on inconsistency synthesis.}
    \vspace{0.6cm}
    \label{tab:fs}
\end{table}

\section{Experiments}
\subsection{Training Strategy}
Our ACDIS comprises four mono-modal encoders $E_{m}$ and one fusion decoder $D_{Fuse}$, accompanied by an additional auxiliary encoder $E_{Aux}$ for knowledge distillation and an auxiliary decoder $D_{Aux}$ for mono-modal segmentation. The architecture of these encoders and decoders closely resembles that of the 3D U-Net. 

We conduct our experiments using the PyTorch framework version 1.13.0. The training process is carried out on two NVIDIA A800-80GB GPUs, spanning 500 epochs, with a batch size of two volumes, consuming in a total 98 GPU hours. We utilize the Adam optimizer with an initial learning rate set to $1\times10^{-4}$ and a weight decay of $1\times10^{-4}$. During the training phase, the input images are randomly cropped to dimensions of $80\times80\times80$. Data augmentation techniques include random rotations, flips, and intensity adjustments.

It is worth noting that our initial training of ACDIS omitted the anatomical constraints, specifically $L_{ACCT}$, as well as the synthesis loss $L_{Syn}$. This is due to the possibility that the auxiliary encoder $E_{Aux}$ may not provide multi-modal features with comprehensive anatomical structures, and the MFSB may encounter difficulties in constructing missing modality features solely based on the limited anatomical information contained within mono-modal features. The anatomical consistency distillation process runs concurrently with our training from the first to the final epoch, while the feature synthesis process starts from the 21st epoch onwards.

\subsection{Datasets and Evaluation Metric}
We preform our experiments on two datasets from the Multi-modal Brain Tumor Segmentation Challenge~\cite{menze2014multimodal}, i.e., BraTS2018 and BraTS2020, which align with that of previous studies~\cite{liu2023m3ae,wang2023multi}. BraTS2018 and BraTS2020 include 285 and 369 cases with ground truth publicly available, respectively. 
For BraTS2020, consistent with~\cite{ding2021rfnet}, we randomly split it into $219:50:100$ for training, validation, and testing, respectively. For BraTS2018, we split it into $199:29:57$, and incorporate a three-fold validation.

Each case of the datasets has four different modalities, i.e., FLAIR, T1ce, T1, and T2 modalities. These modalities are characterized by a volume size of $240\times 240\times 155$, and they capture various properties of brain tumor subregions: GD-enhancing tumor (ET), peritumoral edema (ED), and the necrotic and non-enhancing tumor core (NCR/NET). These subregions of brain tumors are grouped into three nested subregions: the whole tumor (WT), the tumor core (TC), and the enhancing tumor (ET).

Dice coefficient is adopted to evaluate our ACDIS, aligning with~\cite{ding2021rfnet}. The metric is defined as
\begin{align}
    \text{Dice} = \frac{2 \cdot || \bar{y}_{\bar{k}} \cap y_{\bar{k}}||_1}{||\bar{y}_{\bar{k}}||_1 + ||y_{\bar{k}}||_1},
    \label{eq:dice}
\end{align}
where $\bar{k}$ represents different tumor classes, i.e., WT, TC, and ET.

\subsection{Comparisons with the State-of-the-art}
To evaluate the effectiveness of our method, we compare our ACDIS with nine state-of-the-art methods on different cases with missing modalities.The involved methods contain HeMiS~\cite{havaei2016hemis}, U-HVED~\cite{dorent2019hetero}, RFNet~\cite{ding2021rfnet}, UNet-MFI~\cite{zhao2022modality}, mmFormer~\cite{zhang2022mmformer}, MMANet~\cite{wei2023mmanet}, ShaSpec~\cite{wang2023multi}, M3AE~\cite{liu2023m3ae}, and GSS~\cite{qiu2023scratch}. For a fair comparison, all methods are trained under their recommended hyper-parameters within the same dataset split.

As illustrated in Tab.~\ref{tab:main_table}, our method achieves preferable results for most combinations of missing modalities. We achieve improvements of 1.3\%, 1.1\%, and 1.5\% over the second-ranked method on the average DSC for WT, TC, and ET. We also provide a visualization comparison in Figure~\ref{fig:main}, illustrating that our method yields more accurate segmentation results in different combinations of modalities. \textit{More results can be referred to the appendix~\ref{sec:appendix}.}

\subsection{Ablation Study}
We evaluate the proposed components on BraTS2020, employing the average DSC to measure the performance. To maintain the invariance of parameter counts, when removing the AFEB and MFSB blocks, we replace them with $3\times 3\times 3$ convolutions in their respective positions, ensuring a fair comparison.

\paragraph{Effect of consistency distillation.} We conduct experiments within our framework, excluding MFSB to evaluate the effectiveness of consistency distillation. Without AFEB, the model achieves a Dice score of 83.34\% for WT, 75.36\% for TC, and 56.22\% for ET, serving as a baseline for comparison. Then, we cooperate it with multiple distillation losses, including the mean squared error loss ($L_{MSE}$), Kullback-Leibler divergence loss ($L_{KL}$), our proposed variance loss ($L_{Var}$), and our covariance loss ($L_{Covar}$), to constrain the knowledge transferred from multi-modal representations to individual mono-modal representations. As illustrated in Table~\ref{tab:kd}, our ACCT-based approach consistently yielded the most favourable segmentation results. This can be attributed to the specific focus of each loss function. Both $L_{MSE}$ and $L_{KL}$ primarily target adjustments in pixel-wise intensity and intensity distributions, respectively. These adjustments standardize intensity values across different mono-modal features, which can inadvertently suppress modality-specific information that is critical for accurate segmentation. In contrast, our approach enhances segmentation by enriching the local information richness through the variance loss ($L_{Var}$) and promoting anatomical structural similarity through the covariance loss ($L_{Covar}$). These tailored losses improve the consistency between mono-modal and multi-modal features without diminishing the mono-modal inconsistent characteristics. 

\begin{figure}[tbp]
    \centering
    \includegraphics[width=0.98\linewidth]{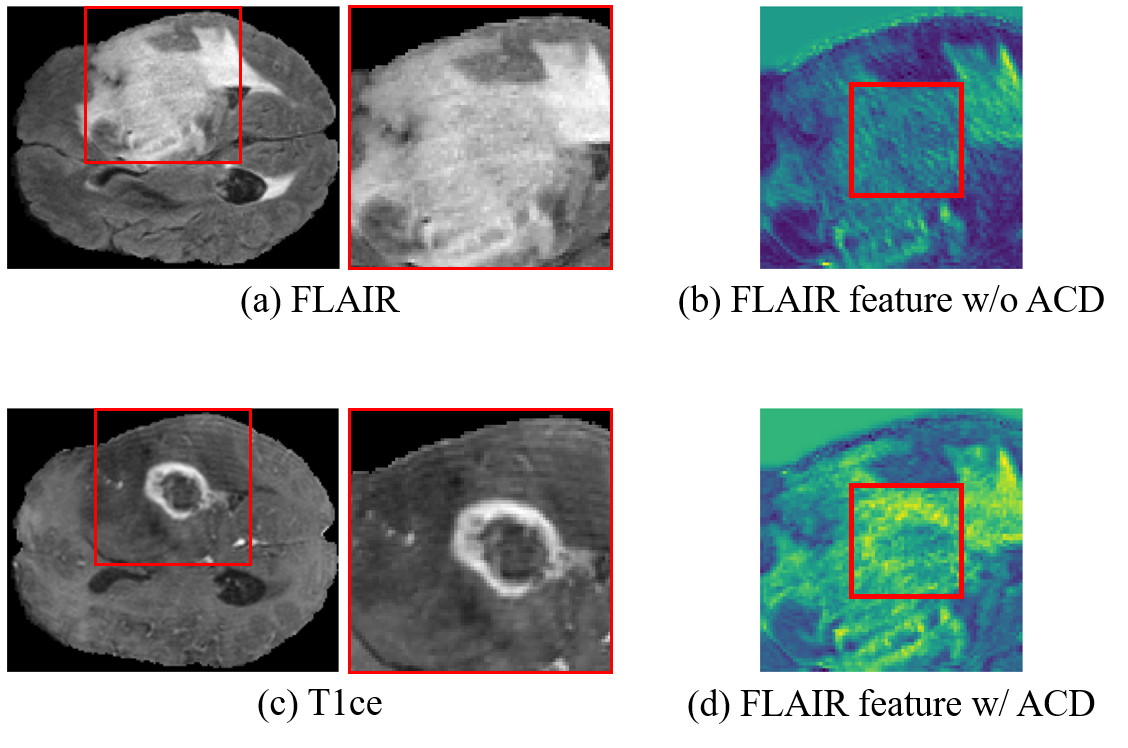}
    \vspace{0.2cm}
    \caption{Feature Visualization: (a) depicts the input image of the FLAIR modality. (b) illustrates the average feature derived from our model without our ACD. (c) showcases the input image of the T1ce modality. Finally, (d) denotes the average feature obtained from our model with ACD, highlighting that the FLAIR features have effectively assimilated characteristics of the T1ce modality through the use of ACD.}
    \vspace{0.6cm}
    \label{fig:acd_vis}
\end{figure}

We provide a visualization in Figure~\ref{fig:acd_vis} to illustrate the effectiveness of our ACD. By comparing Figure~\ref{fig:acd_vis} (b) and Figure~\ref{fig:acd_vis} (d), it is evident that the FLAIR features, with the application of our ACD, are able to capture anatomical structures that are less distinct in the FLAIR modality alone but are prominent in the T1ce modality. This capability demonstrates that, even in the absence of the T1ce modality, our network can successfully extract and highlight these subtle yet consistent anatomical structures from the FLAIR modality alone.

\begin{figure}[tbp]
    \centering
    \includegraphics[width=0.98\linewidth]{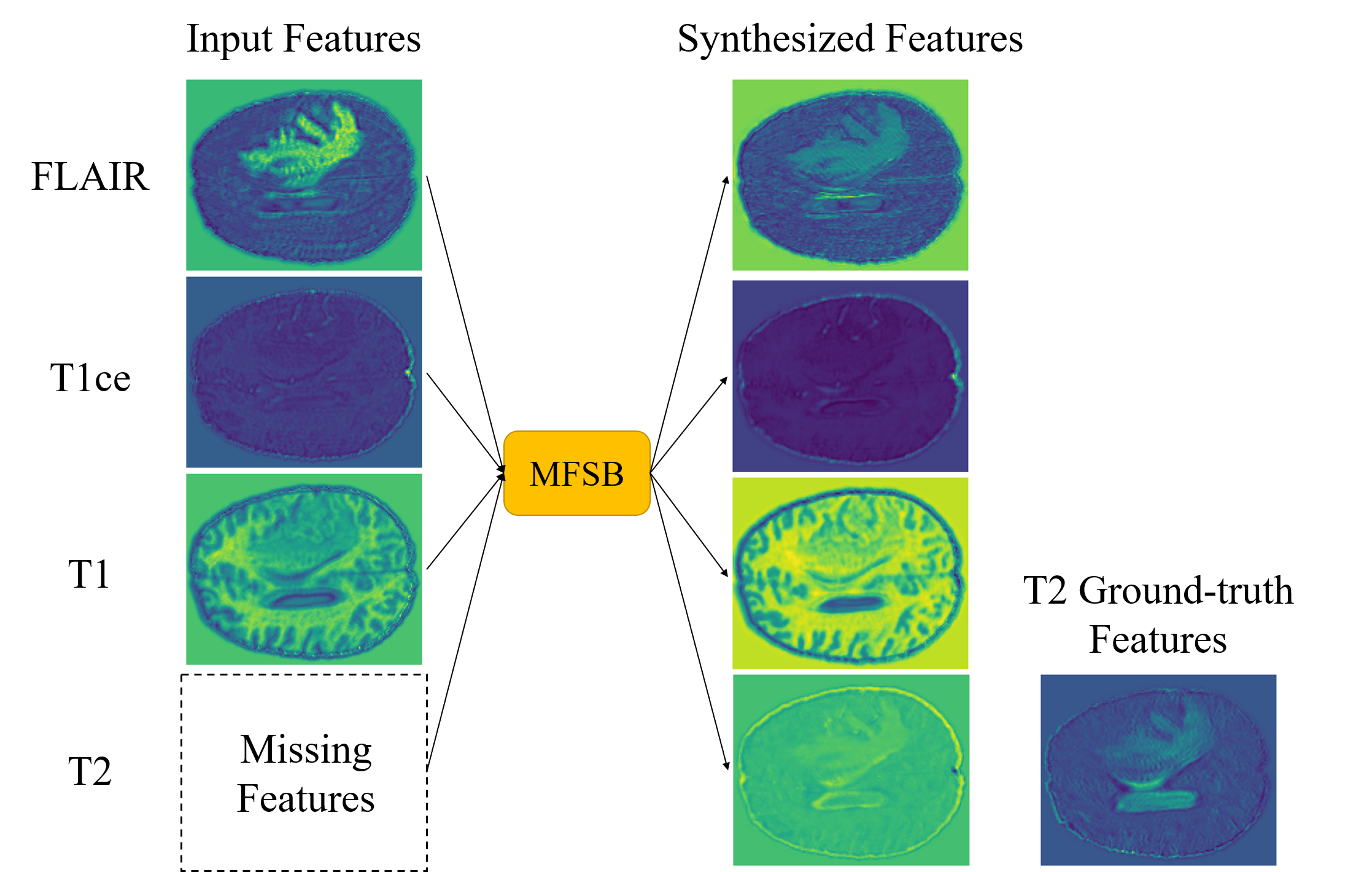}
    \vspace{0.2cm}
    \caption{Feature synthesis process of our MFSB. With the available modalities, containing FLAIR, T1ce, and T1, our MFSB effectively synthesizes the features of the missing modality, i.e., T2 modality. T2 ground-truth features are obtained when complete modalities are provided.}
    \vspace{0.6cm}
    \label{fig:mfsb_vis}
\end{figure}


\paragraph{Effect of inconsistency synthesis.} We further investigate the effectiveness of inconsistency synthesis within our framework, which incorporates ACD. The comparative results are presented in Table~\ref{tab:fs}. In the absence of the Modality Feature Synthesis Block (MFSB), which is designed for synthesizing features of absent modalities, our model achieves Dice scores of 86.61\% for WT, 78.56\% for TC, and 61.09\% for ET. When integrating MFSB alongside other losses, including adversarial loss ($L_{Adver}$), variance loss ($L_{Var}$), covariance loss ($L_{Covar}$), and mean squared error synthesis loss ($L_{Syn}$), our model demonstrates significantly improved performance. Notably, the highest performance is observed when employing $L_{Syn}$, which specifically encourages the MFSB to generate modality-specific features with distinct intensity characteristics that are critical for inconsistency synthesis. However, the combination of $L_{Adver}$ with $L_{Syn}$ results in a performance decline, primarily due to a decrease in data fidelity.

To illustrate our MFSB's feature synthesis process (Figure~\ref{fig:mfsb_vis}), we demonstrate its capability to synthesize missing T2 modality features. FLAIR, T1ce, and T1 modalities undergo encoding via mono-modal encoders and our ACD, resulting in `Input Features' obtained by averaging features across channels. Our MFSB learns modality-specific affine transformations and applies non-linear activation to synthesize missing features (`Synthesized Features'). Remarkably, the synthesized T2 features resemble ground-truth counterparts, distinguished by distinct colors from the other mono-modal features indicating captured anatomical inconsistencies, i.e., modality-specific pixel values.


\section{Conclusion}
In this study, we introduce the Anatomical Consistency Distillation and Inconsistency Synthesis (ACDIS) framework, a novel approach for brain tumor segmentation in cases of missing modalities. ACDIS is composed of two primary components: Anatomical Consistency Distillation (ACD) and the Modality Feature Synthesis Block (MFSB). In ACD, the Anatomical Feature Enhancement Block (AFEB) effectively models the relationship between mono-modal and multi-modal representations, cooperating with Anatomical Consistency ConsTraints (ACCT), which transfer rich multi-modal knowledge and anatomical structural similarities to boost mono-modal representation. Concurrently, MFSB addresses the challenge of inconsistent anatomical information in missing modalities by generating comprehensive modality-specific features from available mono-modal representations. By leveraging both anatomical consistency and inconsistency, ACDIS demonstrates superior segmentation performance for the task of brain tumor segmentation with missing modalities, as validated on BraTS2020 and BraTS2018.

\section{Acknowledgement}
This work was in part supported by the National Natural Science Foundation of China under grants 62032006 and 62021001.



\bibliography{ACDIS}

\onecolumn
\appendix
\section{Appendix}
\label{sec:appendix}
In this section, we present comparisons using BraTS2018. Specifically, we compare our method with five state-of-the-art methods on BraTS2018, employing a three-fold validation approach as detailed in Table~\ref{tab_compare18}. For each method, we provide the mean and standard deviation of the Dice score in varying scenarios with missing modalities. The results consistently indicate that our Anatomical Consistency Distillation and Inconsistency Synthesis (ACDIS) framework outperforms the compared state-of-the-art methods across all tumor categories, including whole tumor (WT), tumor core (TC), and enhancing tumor (ET).

\begin{table*}[h]
    \centering
    \renewcommand\arraystretch{1.5}
    \resizebox{\linewidth}{!}{
        \begin{tabular}{l|l|ccccccccccccccc|c}
            \bottomrule
            \multirow{4}{*}{M}  & FLAIR   & $\bullet$ & $\circ$   & $\circ$   & $\circ$   & $\bullet$ & $\bullet$ & $\bullet$ & $\circ$   & $\circ$   & $\circ$   & $\bullet$ & $\bullet$ & $\bullet$ & $\circ$   & $\bullet$ & \multirow{4}{*}{AVG}   \\
            & T1ce & $\circ$   & $\bullet$ & $\circ$   & $\circ$   & $\bullet$ & $\circ$   & $\circ$   & $\bullet$ & $\bullet$ & $\circ$   & $\bullet$ & $\bullet$ & $\circ$   & $\bullet$ & $\bullet$ &       \\
            & T1  & $\circ$   & $\circ$   & $\bullet$ & $\circ$   & $\circ$   & $\bullet$ & $\circ$   & $\bullet$ & $\circ$   & $\bullet$ & $\bullet$ & $\circ$   & $\bullet$ & $\bullet$ & $\bullet$ &       \\
            & T2  & $\circ$   & $\circ$   & $\circ$   & $\bullet$ & $\circ$   & $\circ$   & $\bullet$ & $\circ$   & $\bullet$ & $\bullet$ & $\circ$   & $\bullet$ & $\bullet$ & $\bullet$ & $\bullet$ &       \\ \hline
            \multirow{15}{*}{\textbf{WT}} 
                                        & UNet-MFI   & \begin{tabular}[c]{@{}c@{}}\cellcolor{lightgray}\underline{84.06}\\${\pm0.24}$\end{tabular} & \begin{tabular}[c]{@{}c@{}}\cellcolor{lightgray}65.90\\${\pm1.73}$\end{tabular} & \begin{tabular}[c]{@{}c@{}}\cellcolor{lightgray}68.80\\${\pm0.95}$\end{tabular} & \begin{tabular}[c]{@{}c@{}}\cellcolor{lightgray}\underline{85.23}\\${\pm0.25}$\end{tabular} & \begin{tabular}[c]{@{}c@{}}\cellcolor{lightgray}90.23\\${\pm0.35}$\end{tabular} & \begin{tabular}[c]{@{}c@{}}\cellcolor{lightgray}\textbf{89.88}\\${\pm0.64}$\end{tabular} & \begin{tabular}[c]{@{}c@{}}\cellcolor{lightgray}\textbf{92.75}\\${\pm0.63}$\end{tabular} & \begin{tabular}[c]{@{}c@{}}\cellcolor{lightgray}75.46\\${\pm0.51}$\end{tabular} & \begin{tabular}[c]{@{}c@{}}\cellcolor{lightgray}88.40\\${\pm0.11}$\end{tabular} & \begin{tabular}[c]{@{}c@{}}\cellcolor{lightgray}\underline{89.47}\\${\pm0.48}$\end{tabular} & \begin{tabular}[c]{@{}c@{}}\cellcolor{lightgray}90.97\\${\pm0.67}$\end{tabular} & \begin{tabular}[c]{@{}c@{}}\cellcolor{lightgray}93.27\\${\pm0.27}$\end{tabular} & \begin{tabular}[c]{@{}c@{}}\cellcolor{lightgray}\underline{93.82}\\${\pm0.33}$\end{tabular} & \begin{tabular}[c]{@{}c@{}}\cellcolor{lightgray}\underline{89.71}\\${\pm0.57}$\end{tabular} & \begin{tabular}[c]{@{}c@{}}\cellcolor{lightgray}\underline{93.81}\\${\pm0.08}$\end{tabular} & \begin{tabular}[c]{@{}c@{}}\cellcolor{lightgray}86.12\\${\pm0.04}$\end{tabular}           \\ 
                                         & RFNet    & \begin{tabular}[c]{@{}c@{}}\cellcolor{lightgray} 83.01\\${\pm3.64}$\end{tabular} & \begin{tabular}[c]{@{}c@{}}\cellcolor{lightgray} 68.94\\${\pm1.11}$\end{tabular} & \begin{tabular}[c]{@{}c@{}}\cellcolor{lightgray} \underline{71.78}\\${\pm1.24}$\end{tabular} & \begin{tabular}[c]{@{}c@{}}\cellcolor{lightgray} 84.27\\${\pm1.13}$\end{tabular} & \begin{tabular}[c]{@{}c@{}}\cellcolor{lightgray} 89.25\\${\pm1.65}$\end{tabular} & \begin{tabular}[c]{@{}c@{}}\cellcolor{lightgray} 89.47\\${\pm1.47}$\end{tabular} & \begin{tabular}[c]{@{}c@{}}\cellcolor{lightgray} 90.50\\${\pm1.79}$\end{tabular} & \begin{tabular}[c]{@{}c@{}}\cellcolor{lightgray} \underline{75.76}\\${\pm0.89}$\end{tabular} & \begin{tabular}[c]{@{}c@{}}\cellcolor{lightgray} 87.30\\${\pm1.21}$\end{tabular} & \begin{tabular}[c]{@{}c@{}}\cellcolor{lightgray} 87.99\\${\pm1.15}$\end{tabular} & \begin{tabular}[c]{@{}c@{}}\cellcolor{lightgray} 90.54\\${\pm1.08}$\end{tabular} & \begin{tabular}[c]{@{}c@{}}\cellcolor{lightgray} 91.57\\${\pm1.28}$\end{tabular} & \begin{tabular}[c]{@{}c@{}}\cellcolor{lightgray} 92.00\\${\pm1.10}$\end{tabular} & \begin{tabular}[c]{@{}c@{}}\cellcolor{lightgray} 88.57\\${\pm1.35}$\end{tabular} & \begin{tabular}[c]{@{}c@{}}\cellcolor{lightgray} 92.18\\${\pm0.99}$\end{tabular} & \begin{tabular}[c]{@{}c@{}}\cellcolor{lightgray} 85.54\\${\pm1.31}$\end{tabular}  \\ 
                                        & mmFormer & \begin{tabular}[c]{@{}c@{}}\cellcolor{lightgray} 81.10\\${\pm0.82}$\end{tabular} & \begin{tabular}[c]{@{}c@{}}\cellcolor{lightgray} 64.44\\${\pm1.18}$\end{tabular} & \begin{tabular}[c]{@{}c@{}}\cellcolor{lightgray} 66.87\\${\pm1.99}$\end{tabular} & \begin{tabular}[c]{@{}c@{}}\cellcolor{lightgray} 82.77\\${\pm0.06}$\end{tabular} & \begin{tabular}[c]{@{}c@{}}\cellcolor{lightgray} 87.22\\${\pm0.84}$\end{tabular} & \begin{tabular}[c]{@{}c@{}}\cellcolor{lightgray} 87.99\\${\pm0.70}$\end{tabular} & \begin{tabular}[c]{@{}c@{}}\cellcolor{lightgray} 89.32\\${\pm0.89}$\end{tabular} & \begin{tabular}[c]{@{}c@{}}\cellcolor{lightgray} 72.33\\${\pm0.66}$\end{tabular} & \begin{tabular}[c]{@{}c@{}}\cellcolor{lightgray} 85.48\\${\pm0.28}$\end{tabular} & \begin{tabular}[c]{@{}c@{}}\cellcolor{lightgray} 86.26\\${\pm0.36}$\end{tabular} & \begin{tabular}[c]{@{}c@{}}\cellcolor{lightgray} 88.28\\${\pm0.65}$\end{tabular} & \begin{tabular}[c]{@{}c@{}}\cellcolor{lightgray} 89.96\\${\pm0.74}$\end{tabular} & \begin{tabular}[c]{@{}c@{}}\cellcolor{lightgray} 90.44\\${\pm0.58}$\end{tabular} & \begin{tabular}[c]{@{}c@{}}\cellcolor{lightgray} 86.61\\${\pm0.50}$\end{tabular} & \begin{tabular}[c]{@{}c@{}}\cellcolor{lightgray} 90.52\\${\pm0.67}$\end{tabular} & \begin{tabular}[c]{@{}c@{}}\cellcolor{lightgray} 83.24\\${\pm0.47}$\end{tabular}   \\
                                        & M3AE & \begin{tabular}[c]{@{}c@{}}\cellcolor{lightgray} 83.48\\${\pm5.25}$\end{tabular} & \begin{tabular}[c]{@{}c@{}}\cellcolor{lightgray} 65.67\\${\pm2.64}$\end{tabular} & \begin{tabular}[c]{@{}c@{}}\cellcolor{lightgray} 70.54\\${\pm2.56}$\end{tabular} & \begin{tabular}[c]{@{}c@{}}\cellcolor{lightgray} 79.68\\${\pm5.13}$\end{tabular} & \begin{tabular}[c]{@{}c@{}}\cellcolor{lightgray} \underline{90.69}\\${\pm3.42}$\end{tabular} & \begin{tabular}[c]{@{}c@{}}\cellcolor{lightgray} 89.39\\${\pm2.65}$\end{tabular} & \begin{tabular}[c]{@{}c@{}}\cellcolor{lightgray} 89.04\\${\pm4.29}$\end{tabular} & \begin{tabular}[c]{@{}c@{}}\cellcolor{lightgray} 75.69\\${\pm1.67}$\end{tabular} & \begin{tabular}[c]{@{}c@{}}\cellcolor{lightgray} 85.64\\${\pm4.56}$\end{tabular} & \begin{tabular}[c]{@{}c@{}}\cellcolor{lightgray} 87.28\\${\pm3.24}$\end{tabular} & \begin{tabular}[c]{@{}c@{}}\cellcolor{lightgray} \textbf{91.35}\\${\pm1.95}$\end{tabular} & \begin{tabular}[c]{@{}c@{}}\cellcolor{lightgray} 92.03\\${\pm3.37}$\end{tabular} & \begin{tabular}[c]{@{}c@{}}\cellcolor{lightgray} 92.55\\${\pm2.89}$\end{tabular} & \begin{tabular}[c]{@{}c@{}}\cellcolor{lightgray} 88.53\\${\pm2.46}$\end{tabular} & \begin{tabular}[c]{@{}c@{}}\cellcolor{lightgray} 92.10\\${\pm2.41}$\end{tabular} & \begin{tabular}[c]{@{}c@{}}\cellcolor{lightgray} 83.98\\${\pm2.94}$\end{tabular}              \\
                                       & ShapSpec & \begin{tabular}[c]{@{}c@{}}\cellcolor{lightgray} 82.76\\${\pm4.05}$\end{tabular} & \begin{tabular}[c]{@{}c@{}}\cellcolor{lightgray} \underline{69.23}\\${\pm2.19}$\end{tabular} & \begin{tabular}[c]{@{}c@{}}\cellcolor{lightgray} 71.09\\${\pm0.75}$\end{tabular} & \begin{tabular}[c]{@{}c@{}}\cellcolor{lightgray} \textbf{85.50}\\${\pm1.02}$\end{tabular} & \begin{tabular}[c]{@{}c@{}}\cellcolor{lightgray} 90.43\\${\pm1.96}$\end{tabular} & \begin{tabular}[c]{@{}c@{}}\cellcolor{lightgray} 89.09\\${\pm1.30}$\end{tabular} & \begin{tabular}[c]{@{}c@{}}\cellcolor{lightgray} \underline{92.01}\\${\pm1.71}$\end{tabular} & \begin{tabular}[c]{@{}c@{}}\cellcolor{lightgray} 75.74\\${\pm1.25}$\end{tabular} & \begin{tabular}[c]{@{}c@{}}\cellcolor{lightgray} \textbf{88.72}\\${\pm1.09}$\end{tabular} & \begin{tabular}[c]{@{}c@{}}\cellcolor{lightgray} 89.06\\${\pm0.69}$\end{tabular} & \begin{tabular}[c]{@{}c@{}}\cellcolor{lightgray} \underline{91.37}\\${\pm1.32}$\end{tabular} & \begin{tabular}[c]{@{}c@{}}\cellcolor{lightgray} \underline{93.03}\\${\pm1.29}$\end{tabular} & \begin{tabular}[c]{@{}c@{}}\cellcolor{lightgray} 93.27\\${\pm0.95}$\end{tabular} & \begin{tabular}[c]{@{}c@{}}\cellcolor{lightgray} \underline{89.71}\\${\pm1.11}$\end{tabular} & \begin{tabular}[c]{@{}c@{}}\cellcolor{lightgray} 93.39\\${\pm1.03}$\end{tabular} & \begin{tabular}[c]{@{}c@{}}\cellcolor{lightgray} \underline{86.36}\\${\pm1.34}$\end{tabular}  \\       
                                        & Ours    & \begin{tabular}[c]{@{}c@{}}\cellcolor{lightgray} \textbf{85.12}\\${\pm0.72}$\end{tabular} & \begin{tabular}[c]{@{}c@{}}\cellcolor{lightgray} \textbf{73.72}\\${\pm0.93}$\end{tabular} & \begin{tabular}[c]{@{}c@{}}\cellcolor{lightgray} \textbf{74.81}\\${\pm0.67}$\end{tabular} & \begin{tabular}[c]{@{}c@{}}\cellcolor{lightgray} 84.83\\${\pm0.62}$\end{tabular} & \begin{tabular}[c]{@{}c@{}}\cellcolor{lightgray} \textbf{93.17}\\${\pm0.67}$\end{tabular} & \begin{tabular}[c]{@{}c@{}}\cellcolor{lightgray} \underline{89.83}\\${\pm0.72}$\end{tabular} & \begin{tabular}[c]{@{}c@{}}\cellcolor{lightgray} 90.03\\${\pm0.76}$\end{tabular} & \begin{tabular}[c]{@{}c@{}}\cellcolor{lightgray} \textbf{80.84}\\${\pm0.73}$\end{tabular} & \begin{tabular}[c]{@{}c@{}}\cellcolor{lightgray} 88.27\\${\pm0.70}$\end{tabular} & \begin{tabular}[c]{@{}c@{}}\cellcolor{lightgray} \textbf{91.60}\\${\pm0.69}$\end{tabular} & \begin{tabular}[c]{@{}c@{}}\cellcolor{lightgray} 91.24\\${\pm0.71}$\end{tabular} & \begin{tabular}[c]{@{}c@{}}\cellcolor{lightgray} \textbf{93.85}\\${\pm0.74}$\end{tabular} & \begin{tabular}[c]{@{}c@{}}\cellcolor{lightgray} \textbf{94.41}\\${\pm0.76}$\end{tabular} & \begin{tabular}[c]{@{}c@{}}\cellcolor{lightgray} \textbf{92.48}\\${\pm0.75}$\end{tabular} & \begin{tabular}[c]{@{}c@{}}\cellcolor{lightgray} \textbf{95.89}\\${\pm0.77}$\end{tabular} & \begin{tabular}[c]{@{}c@{}}\cellcolor{lightgray} \textbf{88.01}\\${\pm0.72}$\end{tabular}  \\ \hline
                        \multirow{15}{*}{\textbf{TC}} 
                                        & UNet-MFI   & \begin{tabular}[c]{@{}c@{}}\cellcolor{lightgray}56.77\\${\pm1.25}$\end{tabular} & \begin{tabular}[c]{@{}c@{}}\cellcolor{lightgray}53.93\\${\pm1.55}$\end{tabular} & \begin{tabular}[c]{@{}c@{}}\cellcolor{lightgray}\underline{80.70}\\${\pm0.52}$\end{tabular} & \begin{tabular}[c]{@{}c@{}}\cellcolor{lightgray}63.35\\${\pm1.21}$\end{tabular} & \begin{tabular}[c]{@{}c@{}}\cellcolor{lightgray}69.62\\${\pm0.05}$\end{tabular} & \begin{tabular}[c]{@{}c@{}}\cellcolor{lightgray}84.89\\${\pm0.47}$\end{tabular} & \begin{tabular}[c]{@{}c@{}}\cellcolor{lightgray}69.86\\${\pm0.45}$\end{tabular} & \begin{tabular}[c]{@{}c@{}}\cellcolor{lightgray}\underline{85.11}\\${\pm0.37}$\end{tabular} & \begin{tabular}[c]{@{}c@{}}\cellcolor{lightgray}68.82\\${\pm1.65}$\end{tabular} & \begin{tabular}[c]{@{}c@{}}\cellcolor{lightgray}87.24\\${\pm0.67}$\end{tabular} & \begin{tabular}[c]{@{}c@{}}\cellcolor{lightgray}88.15\\${\pm0.16}$\end{tabular} & \begin{tabular}[c]{@{}c@{}}\cellcolor{lightgray}73.36\\${\pm0.61}$\end{tabular} & \begin{tabular}[c]{@{}c@{}}\cellcolor{lightgray}89.13\\${\pm0.39}$\end{tabular} & \begin{tabular}[c]{@{}c@{}}\cellcolor{lightgray}\underline{89.27}\\${\pm0.97}$\end{tabular} & \begin{tabular}[c]{@{}c@{}}\cellcolor{lightgray}90.17\\${\pm0.03}$\end{tabular} & \begin{tabular}[c]{@{}c@{}}\cellcolor{lightgray}76.69\\${\pm0.51}$\end{tabular}  \\ 
                                        & RFNet    & \begin{tabular}[c]{@{}c@{}}\cellcolor{lightgray} 60.88\\${\pm3.43}$\end{tabular} & \begin{tabular}[c]{@{}c@{}}\cellcolor{lightgray} 57.42\\${\pm1.48}$\end{tabular} & \begin{tabular}[c]{@{}c@{}}\cellcolor{lightgray} 77.81\\${\pm1.84}$\end{tabular} & \begin{tabular}[c]{@{}c@{}}\cellcolor{lightgray} \underline{63.89}\\${\pm3.31}$\end{tabular} & \begin{tabular}[c]{@{}c@{}}\cellcolor{lightgray} 71.58\\${\pm1.82}$\end{tabular} & \begin{tabular}[c]{@{}c@{}}\cellcolor{lightgray} 85.99\\${\pm1.71}$\end{tabular} & \begin{tabular}[c]{@{}c@{}}\cellcolor{lightgray} 71.40\\${\pm1.51}$\end{tabular} & \begin{tabular}[c]{@{}c@{}}\cellcolor{lightgray} 80.37\\${\pm1.53}$\end{tabular} & \begin{tabular}[c]{@{}c@{}}\cellcolor{lightgray} 70.21\\${\pm1.41}$\end{tabular} & \begin{tabular}[c]{@{}c@{}}\cellcolor{lightgray} 85.83\\${\pm1.33}$\end{tabular} & \begin{tabular}[c]{@{}c@{}}\cellcolor{lightgray} 86.13\\${\pm1.52}$\end{tabular} & \begin{tabular}[c]{@{}c@{}}\cellcolor{lightgray} 74.21\\${\pm0.78}$\end{tabular} & \begin{tabular}[c]{@{}c@{}}\cellcolor{lightgray} 87.56\\${\pm1.40}$\end{tabular} & \begin{tabular}[c]{@{}c@{}}\cellcolor{lightgray} 86.20\\${\pm1.53}$\end{tabular} & \begin{tabular}[c]{@{}c@{}}\cellcolor{lightgray} 87.90\\${\pm1.50}$\end{tabular} & \begin{tabular}[c]{@{}c@{}}\cellcolor{lightgray} 76.43\\${\pm1.19}$\end{tabular} \\ 
                                        & mmFormer & \begin{tabular}[c]{@{}c@{}}\cellcolor{lightgray} 52.60\\${\pm2.05}$\end{tabular} & \begin{tabular}[c]{@{}c@{}}\cellcolor{lightgray} 52.69\\${\pm0.59}$\end{tabular} & \begin{tabular}[c]{@{}c@{}}\cellcolor{lightgray} 77.61\\${\pm0.13}$\end{tabular} & \begin{tabular}[c]{@{}c@{}}\cellcolor{lightgray} 61.18\\${\pm2.56}$\end{tabular} & \begin{tabular}[c]{@{}c@{}}\cellcolor{lightgray} 66.22\\${\pm0.54}$\end{tabular} & \begin{tabular}[c]{@{}c@{}}\cellcolor{lightgray} 82.07\\${\pm0.29}$\end{tabular} & \begin{tabular}[c]{@{}c@{}}\cellcolor{lightgray} 66.25\\${\pm2.34}$\end{tabular} & \begin{tabular}[c]{@{}c@{}}\cellcolor{lightgray} 81.10\\${\pm0.63}$\end{tabular} & \begin{tabular}[c]{@{}c@{}}\cellcolor{lightgray} 65.87\\${\pm2.07}$\end{tabular} & \begin{tabular}[c]{@{}c@{}}\cellcolor{lightgray} 84.20\\${\pm1.12}$\end{tabular} & \begin{tabular}[c]{@{}c@{}}\cellcolor{lightgray} 84.79\\${\pm0.12}$\end{tabular} & \begin{tabular}[c]{@{}c@{}}\cellcolor{lightgray} 69.58\\${\pm1.04}$\end{tabular} & \begin{tabular}[c]{@{}c@{}}\cellcolor{lightgray} 85.78\\${\pm0.69}$\end{tabular} & \begin{tabular}[c]{@{}c@{}}\cellcolor{lightgray} 85.63\\${\pm0.81}$\end{tabular} & \begin{tabular}[c]{@{}c@{}}\cellcolor{lightgray} 86.47\\${\pm0.14}$\end{tabular} & \begin{tabular}[c]{@{}c@{}}\cellcolor{lightgray} 73.47\\${\pm0.60}$\end{tabular}  \\
                                        & M3AE & \begin{tabular}[c]{@{}c@{}}\cellcolor{lightgray} \underline{61.45}\\${\pm2.35}$\end{tabular} & \begin{tabular}[c]{@{}c@{}}\cellcolor{lightgray} 56.93\\${\pm1.88}$\end{tabular} & \begin{tabular}[c]{@{}c@{}}\cellcolor{lightgray} 77.97\\${\pm3.82}$\end{tabular} & \begin{tabular}[c]{@{}c@{}}\cellcolor{lightgray} 61.85\\${\pm1.57}$\end{tabular} & \begin{tabular}[c]{@{}c@{}}\cellcolor{lightgray} 73.08\\${\pm2.07}$\end{tabular} & \begin{tabular}[c]{@{}c@{}}\cellcolor{lightgray} 86.50\\${\pm2.69}$\end{tabular} & \begin{tabular}[c]{@{}c@{}}\cellcolor{lightgray} 71.45\\${\pm2.73}$\end{tabular} & \begin{tabular}[c]{@{}c@{}}\cellcolor{lightgray} 81.48\\${\pm3.26}$\end{tabular} & \begin{tabular}[c]{@{}c@{}}\cellcolor{lightgray} 69.44\\${\pm1.66}$\end{tabular} & \begin{tabular}[c]{@{}c@{}}\cellcolor{lightgray} \underline{87.53}\\${\pm2.47}$\end{tabular} & \begin{tabular}[c]{@{}c@{}}\cellcolor{lightgray} \underline{88.66}\\${\pm2.24}$\end{tabular} & \begin{tabular}[c]{@{}c@{}}\cellcolor{lightgray} 75.35\\${\pm1.81}$\end{tabular} & \begin{tabular}[c]{@{}c@{}}\cellcolor{lightgray} \textbf{90.47}\\${\pm2.17}$\end{tabular} & \begin{tabular}[c]{@{}c@{}}\cellcolor{lightgray} 88.45\\${\pm2.33}$\end{tabular} & \begin{tabular}[c]{@{}c@{}}\cellcolor{lightgray} \underline{90.91}\\${\pm2.13}$\end{tabular} & \begin{tabular}[c]{@{}c@{}}\cellcolor{lightgray} 77.37\\${\pm1.61}$\end{tabular} \\
                                        & ShapSpec & \begin{tabular}[c]{@{}c@{}}\cellcolor{lightgray} 60.71\\${\pm4.03}$\end{tabular} & \begin{tabular}[c]{@{}c@{}}\cellcolor{lightgray} \underline{58.56}\\${\pm1.10}$\end{tabular} & \begin{tabular}[c]{@{}c@{}}\cellcolor{lightgray} 78.22\\${\pm1.57}$\end{tabular} & \begin{tabular}[c]{@{}c@{}}\cellcolor{lightgray} \textbf{64.80}\\${\pm3.40}$\end{tabular} & \begin{tabular}[c]{@{}c@{}}\cellcolor{lightgray} \underline{73.34}\\${\pm2.20}$\end{tabular} & \begin{tabular}[c]{@{}c@{}}\cellcolor{lightgray} \underline{86.65}\\${\pm1.59}$\end{tabular} & \begin{tabular}[c]{@{}c@{}}\cellcolor{lightgray} \underline{72.58}\\${\pm2.25}$\end{tabular} & \begin{tabular}[c]{@{}c@{}}\cellcolor{lightgray} 81.72\\${\pm1.39}$\end{tabular} & \begin{tabular}[c]{@{}c@{}}\cellcolor{lightgray} \underline{70.81}\\${\pm1.91}$\end{tabular} & \begin{tabular}[c]{@{}c@{}}\cellcolor{lightgray} \textbf{87.54}\\${\pm1.24}$\end{tabular} & \begin{tabular}[c]{@{}c@{}}\cellcolor{lightgray} 88.03\\${\pm1.68}$\end{tabular} & \begin{tabular}[c]{@{}c@{}}\cellcolor{lightgray} \underline{75.77}\\${\pm1.23}$\end{tabular} & \begin{tabular}[c]{@{}c@{}}\cellcolor{lightgray} 89.88\\${\pm1.30}$\end{tabular} & \begin{tabular}[c]{@{}c@{}}\cellcolor{lightgray} 87.99\\${\pm1.39}$\end{tabular} & \begin{tabular}[c]{@{}c@{}}\cellcolor{lightgray} 90.26\\${\pm1.42}$\end{tabular} & \begin{tabular}[c]{@{}c@{}}\cellcolor{lightgray} \underline{77.79}\\${\pm1.15}$\end{tabular} \\
                                        & Ours  & \begin{tabular}[c]{@{}c@{}}\cellcolor{lightgray} \textbf{63.29}\\${\pm1.77}$\end{tabular} & \begin{tabular}[c]{@{}c@{}}\cellcolor{lightgray} \textbf{60.15}\\${\pm1.99}$\end{tabular} & \begin{tabular}[c]{@{}c@{}}\cellcolor{lightgray} \textbf{81.76}\\${\pm2.01}$\end{tabular} & \begin{tabular}[c]{@{}c@{}}\cellcolor{lightgray} 63.00\\${\pm1.88}$\end{tabular} & \begin{tabular}[c]{@{}c@{}}\cellcolor{lightgray} \textbf{75.62}\\${\pm1.87}$\end{tabular} & \begin{tabular}[c]{@{}c@{}}\cellcolor{lightgray} \textbf{87.20}\\${\pm1.88}$\end{tabular} & \begin{tabular}[c]{@{}c@{}}\cellcolor{lightgray} \textbf{73.16}\\${\pm1.87}$\end{tabular} & \begin{tabular}[c]{@{}c@{}}\cellcolor{lightgray} \textbf{85.59}\\${\pm1.91}$\end{tabular} & \begin{tabular}[c]{@{}c@{}}\cellcolor{lightgray} \textbf{71.26}\\${\pm1.85}$\end{tabular} & \begin{tabular}[c]{@{}c@{}}\cellcolor{lightgray} 86.49\\${\pm1.85}$\end{tabular} & \begin{tabular}[c]{@{}c@{}}\cellcolor{lightgray} \textbf{89.55}\\${\pm1.86}$\end{tabular} & \begin{tabular}[c]{@{}c@{}}\cellcolor{lightgray} \textbf{79.39}\\${\pm1.87}$\end{tabular} & \begin{tabular}[c]{@{}c@{}}\cellcolor{lightgray} \underline{90.23}\\${\pm1.84}$\end{tabular} & \begin{tabular}[c]{@{}c@{}}\cellcolor{lightgray} \textbf{89.90}\\${\pm1.84}$\end{tabular} & \begin{tabular}[c]{@{}c@{}}\cellcolor{lightgray} \textbf{91.55}\\${\pm1.87}$\end{tabular} & \begin{tabular}[c]{@{}c@{}}\cellcolor{lightgray} \textbf{79.28}\\${\pm1.87}$\end{tabular} \\ \hline
            \multirow{15}{*}{\textbf{ET}} 
                                        & UNet-MFI   & \begin{tabular}[c]{@{}c@{}}\cellcolor{lightgray}37.69\\${\pm1.14}$\end{tabular} & \begin{tabular}[c]{@{}c@{}}\cellcolor{lightgray}25.23\\${\pm4.79}$\end{tabular} & \begin{tabular}[c]{@{}c@{}}\cellcolor{lightgray}74.22\\${\pm1.73}$\end{tabular} & \begin{tabular}[c]{@{}c@{}}\cellcolor{lightgray}41.37\\${\pm1.62}$\end{tabular} & \begin{tabular}[c]{@{}c@{}}\cellcolor{lightgray}41.27\\${\pm1.61}$\end{tabular} & \begin{tabular}[c]{@{}c@{}}\cellcolor{lightgray}\underline{77.88}\\${\pm1.20}$\end{tabular} & \begin{tabular}[c]{@{}c@{}}\cellcolor{lightgray}44.86\\${\pm1.84}$\end{tabular} & \begin{tabular}[c]{@{}c@{}}\cellcolor{lightgray}\underline{76.67}\\${\pm1.23}$\end{tabular} & \begin{tabular}[c]{@{}c@{}}\cellcolor{lightgray}42.87\\${\pm1.57}$\end{tabular} & \begin{tabular}[c]{@{}c@{}}\cellcolor{lightgray}\underline{78.48}\\${\pm0.38}$\end{tabular} & \begin{tabular}[c]{@{}c@{}}\cellcolor{lightgray}78.04\\${\pm1.06}$\end{tabular} & \begin{tabular}[c]{@{}c@{}}\cellcolor{lightgray}46.42\\${\pm1.51}$\end{tabular} & \begin{tabular}[c]{@{}c@{}}\cellcolor{lightgray}78.41\\${\pm0.50}$\end{tabular} & \begin{tabular}[c]{@{}c@{}}\cellcolor{lightgray}78.98\\${\pm0.30}$\end{tabular} & \begin{tabular}[c]{@{}c@{}}\cellcolor{lightgray}78.96\\${\pm0.74}$\end{tabular} & \begin{tabular}[c]{@{}c@{}}\cellcolor{lightgray}60.16\\${\pm0.46}$\end{tabular} \\ 
                                        & RFNet    & \begin{tabular}[c]{@{}c@{}}\cellcolor{lightgray} 37.21\\${\pm1.44}$\end{tabular} & \begin{tabular}[c]{@{}c@{}}\cellcolor{lightgray} 31.97\\${\pm1.04}$\end{tabular} & \begin{tabular}[c]{@{}c@{}}\cellcolor{lightgray} 66.43\\${\pm2.31}$\end{tabular} & \begin{tabular}[c]{@{}c@{}}\cellcolor{lightgray} 40.72\\${\pm1.80}$\end{tabular} & \begin{tabular}[c]{@{}c@{}}\cellcolor{lightgray} 42.30\\${\pm1.65}$\end{tabular} & \begin{tabular}[c]{@{}c@{}}\cellcolor{lightgray} 71.54\\${\pm1.08}$\end{tabular} & \begin{tabular}[c]{@{}c@{}}\cellcolor{lightgray} 44.12\\${\pm0.39}$\end{tabular} & \begin{tabular}[c]{@{}c@{}}\cellcolor{lightgray} 68.96\\${\pm1.31}$\end{tabular} & \begin{tabular}[c]{@{}c@{}}\cellcolor{lightgray} 43.37\\${\pm0.79}$\end{tabular} & \begin{tabular}[c]{@{}c@{}}\cellcolor{lightgray} 72.19\\${\pm1.35}$\end{tabular} & \begin{tabular}[c]{@{}c@{}}\cellcolor{lightgray} 72.33\\${\pm1.41}$\end{tabular} & \begin{tabular}[c]{@{}c@{}}\cellcolor{lightgray} 45.54\\${\pm0.54}$\end{tabular} & \begin{tabular}[c]{@{}c@{}}\cellcolor{lightgray} 73.24\\${\pm1.45}$\end{tabular} & \begin{tabular}[c]{@{}c@{}}\cellcolor{lightgray} 72.53\\${\pm1.46}$\end{tabular} & \begin{tabular}[c]{@{}c@{}}\cellcolor{lightgray} 73.49\\${\pm1.44}$\end{tabular} & \begin{tabular}[c]{@{}c@{}}\cellcolor{lightgray} 57.06\\${\pm0.89}$\end{tabular} \\ 
                                        & mmFormer & \begin{tabular}[c]{@{}c@{}}\cellcolor{lightgray} 32.22\\${\pm1.77}$\end{tabular} & \begin{tabular}[c]{@{}c@{}}\cellcolor{lightgray} 24.93\\${\pm1.98}$\end{tabular} & \begin{tabular}[c]{@{}c@{}}\cellcolor{lightgray} \textbf{75.80}\\${\pm0.78}$\end{tabular} & \begin{tabular}[c]{@{}c@{}}\cellcolor{lightgray} 41.27\\${\pm1.63}$\end{tabular} & \begin{tabular}[c]{@{}c@{}}\cellcolor{lightgray} 39.69\\${\pm1.23}$\end{tabular} & \begin{tabular}[c]{@{}c@{}}\cellcolor{lightgray} \textbf{79.10}\\${\pm0.75}$\end{tabular} & \begin{tabular}[c]{@{}c@{}}\cellcolor{lightgray} 44.31\\${\pm2.03}$\end{tabular} & \begin{tabular}[c]{@{}c@{}}\cellcolor{lightgray} \textbf{78.20}\\${\pm0.52}$\end{tabular} & \begin{tabular}[c]{@{}c@{}}\cellcolor{lightgray} 43.04\\${\pm1.97}$\end{tabular} & \begin{tabular}[c]{@{}c@{}}\cellcolor{lightgray} 79.51\\${\pm0.21}$\end{tabular} & \begin{tabular}[c]{@{}c@{}}\cellcolor{lightgray} \underline{80.08}\\${\pm0.33}$\end{tabular} & \begin{tabular}[c]{@{}c@{}}\cellcolor{lightgray} 45.69\\${\pm1.20}$\end{tabular} & \begin{tabular}[c]{@{}c@{}}\cellcolor{lightgray} \textbf{79.38}\\${\pm0.38}$\end{tabular} & \begin{tabular}[c]{@{}c@{}}\cellcolor{lightgray} \textbf{80.24}\\${\pm0.07}$\end{tabular} & \begin{tabular}[c]{@{}c@{}}\cellcolor{lightgray} \underline{80.06}\\${\pm0.50}$\end{tabular} & \begin{tabular}[c]{@{}c@{}}\cellcolor{lightgray} 60.23\\${\pm0.49}$\end{tabular} \\
                                        & M3AE & \begin{tabular}[c]{@{}c@{}}\cellcolor{lightgray} \underline{40.50}\\${\pm2.34}$\end{tabular} & \begin{tabular}[c]{@{}c@{}}\cellcolor{lightgray} 33.50\\${\pm3.65}$\end{tabular} & \begin{tabular}[c]{@{}c@{}}\cellcolor{lightgray} 68.13\\${\pm4.13}$\end{tabular} & \begin{tabular}[c]{@{}c@{}}\cellcolor{lightgray} 41.40\\${\pm1.35}$\end{tabular} & \begin{tabular}[c]{@{}c@{}}\cellcolor{lightgray} \underline{46.40}\\${\pm3.02}$\end{tabular} & \begin{tabular}[c]{@{}c@{}}\cellcolor{lightgray} 75.28\\${\pm3.09}$\end{tabular} & \begin{tabular}[c]{@{}c@{}}\cellcolor{lightgray} 46.62\\${\pm1.06}$\end{tabular} & \begin{tabular}[c]{@{}c@{}}\cellcolor{lightgray} 71.51\\${\pm3.55}$\end{tabular} & \begin{tabular}[c]{@{}c@{}}\cellcolor{lightgray} 44.55\\${\pm0.89}$\end{tabular} & \begin{tabular}[c]{@{}c@{}}\cellcolor{lightgray} 75.04\\${\pm3.22}$\end{tabular} & \begin{tabular}[c]{@{}c@{}}\cellcolor{lightgray} 76.32\\${\pm2.82}$\end{tabular} & \begin{tabular}[c]{@{}c@{}}\cellcolor{lightgray} 48.65\\${\pm1.19}$\end{tabular} & \begin{tabular}[c]{@{}c@{}}\cellcolor{lightgray} 77.33\\${\pm2.50}$\end{tabular} & \begin{tabular}[c]{@{}c@{}}\cellcolor{lightgray} 76.87\\${\pm2.81}$\end{tabular} & \begin{tabular}[c]{@{}c@{}}\cellcolor{lightgray} 77.60\\${\pm2.68}$\end{tabular} & \begin{tabular}[c]{@{}c@{}}\cellcolor{lightgray} 60.96\\${\pm2.30}$\end{tabular} \\
                                        & ShapSpec & \begin{tabular}[c]{@{}c@{}}\cellcolor{lightgray} 39.73\\${\pm0.44}$\end{tabular} & \begin{tabular}[c]{@{}c@{}}\cellcolor{lightgray} \underline{35.26}\\${\pm1.08}$\end{tabular} & \begin{tabular}[c]{@{}c@{}}\cellcolor{lightgray} \underline{69.94}\\${\pm2.24}$\end{tabular} & \begin{tabular}[c]{@{}c@{}}\cellcolor{lightgray} \underline{44.40}\\${\pm2.46}$\end{tabular} & \begin{tabular}[c]{@{}c@{}}\cellcolor{lightgray} 45.83\\${\pm0.44}$\end{tabular} & \begin{tabular}[c]{@{}c@{}}\cellcolor{lightgray} \underline{76.05}\\${\pm1.30}$\end{tabular} & \begin{tabular}[c]{@{}c@{}}\cellcolor{lightgray} \underline{47.81}\\${\pm1.52}$\end{tabular} & \begin{tabular}[c]{@{}c@{}}\cellcolor{lightgray} 73.16\\${\pm1.88}$\end{tabular} & \begin{tabular}[c]{@{}c@{}}\cellcolor{lightgray} \underline{46.00}\\${\pm1.78}$\end{tabular} & \begin{tabular}[c]{@{}c@{}}\cellcolor{lightgray} \underline{76.78}\\${\pm1.55}$\end{tabular} & \begin{tabular}[c]{@{}c@{}}\cellcolor{lightgray} 77.10\\${\pm1.56}$\end{tabular} & \begin{tabular}[c]{@{}c@{}}\cellcolor{lightgray} \underline{49.45}\\${\pm0.40}$\end{tabular} & \begin{tabular}[c]{@{}c@{}}\cellcolor{lightgray} \underline{78.18}\\${\pm1.47}$\end{tabular} & \begin{tabular}[c]{@{}c@{}}\cellcolor{lightgray} \underline{77.24}\\${\pm1.61}$\end{tabular} & \begin{tabular}[c]{@{}c@{}}\cellcolor{lightgray} 78.35\\${\pm1.48}$\end{tabular} & \begin{tabular}[c]{@{}c@{}}\cellcolor{lightgray} \underline{61.09}\\${\pm0.42}$\end{tabular} \\
                                        & Ours   & \begin{tabular}[c]{@{}c@{}}\cellcolor{lightgray} \textbf{41.54}\\${\pm1.44}$\end{tabular} & \begin{tabular}[c]{@{}c@{}}\cellcolor{lightgray} \textbf{36.42}\\${\pm0.76}$\end{tabular} & \begin{tabular}[c]{@{}c@{}}\cellcolor{lightgray} 68.81\\${\pm1.08}$\end{tabular} & \begin{tabular}[c]{@{}c@{}}\cellcolor{lightgray} \textbf{45.56}\\${\pm1.29}$\end{tabular} & \begin{tabular}[c]{@{}c@{}}\cellcolor{lightgray}\cellcolor{lightgray} \textbf{47.94}\\${\pm1.39}$\end{tabular} & \begin{tabular}[c]{@{}c@{}}\cellcolor{lightgray} 76.59\\${\pm1.43}$\end{tabular} & \begin{tabular}[c]{@{}c@{}}\cellcolor{lightgray} \textbf{52.56}\\${\pm1.42}$\end{tabular} & \begin{tabular}[c]{@{}c@{}}\cellcolor{lightgray} 75.75\\${\pm1.42}$\end{tabular} & \begin{tabular}[c]{@{}c@{}}\cellcolor{lightgray} \textbf{48.59}\\${\pm1.53}$\end{tabular} & \begin{tabular}[c]{@{}c@{}}\cellcolor{lightgray} \textbf{78.92}\\${\pm1.54}$\end{tabular} & \begin{tabular}[c]{@{}c@{}}\cellcolor{lightgray} \textbf{81.86}\\${\pm1.57}$\end{tabular} & \begin{tabular}[c]{@{}c@{}}\cellcolor{lightgray} \textbf{53.96}\\${\pm1.56}$\end{tabular} & \begin{tabular}[c]{@{}c@{}}\cellcolor{lightgray} \underline{78.45}\\${\pm1.56}$\end{tabular} & \begin{tabular}[c]{@{}c@{}}\cellcolor{lightgray} \underline{79.81}\\${\pm1.57}$\end{tabular} & \begin{tabular}[c]{@{}c@{}}\cellcolor{lightgray} \textbf{80.96}\\${\pm1.41}$\end{tabular} & \begin{tabular}[c]{@{}c@{}}\cellcolor{lightgray} \textbf{63.31}\\${\pm1.40}$\end{tabular} \\ \toprule           
            \end{tabular}}
            \vspace{0.2cm}
        \caption{Comparisons with five SOTA methods, including RFNet, UNet-MFI, mmFormer, M3AE, and ShaSpec on BraTS2018. The performance on whole tumor, tumor core, and enhancing tumor segmentation are evaluated by the dice scores (reported as Mean ± Standard Deviation). \textbf{Bold} and \underline{underlined} indicate the $1st$ and $2nd$ ranks, respectively. Additionally, we use • to denote available modalities and ◦ to denote missing modalities.}
        \vspace{0.6cm}
        \label{tab_compare18}
    \end{table*}

\end{document}